\documentclass[a4paper]{article}
\usepackage{lmodern}
\usepackage{amssymb,amsmath}
\usepackage{ifxetex,ifluatex}
\usepackage{fixltx2e} 
\ifnum 0\ifxetex 1\fi\ifluatex 1\fi=0 
  \usepackage[T1]{fontenc}
  \usepackage[utf8]{inputenc}
\else 
  \ifxetex
    \usepackage{mathspec}
  \else
    \usepackage{fontspec}
  \fi
  \defaultfontfeatures{Ligatures=TeX,Scale=MatchLowercase}
\fi
\IfFileExists{upquote.sty}{\usepackage{upquote}}{}
\IfFileExists{microtype.sty}{%
\usepackage{microtype}
\UseMicrotypeSet[protrusion]{basicmath} 
}{}
\usepackage[margin=1in]{geometry}
\usepackage{hyperref}
\hypersetup{unicode=true,
            pdfborder={0 0 0},
            breaklinks=true}
\usepackage{longtable,booktabs}
\usepackage{graphicx,grffile}
\makeatletter
\def\maxwidth{\ifdim\Gin@nat@width>\linewidth\linewidth\else\Gin@nat@width\fi}
\def\maxheight{\ifdim\Gin@nat@height>\textheight\textheight\else\Gin@nat@height\fi}
\makeatother
\setkeys{Gin}{width=\maxwidth,height=\maxheight,keepaspectratio}
\IfFileExists{parskip.sty}{%
\usepackage{parskip}
}{
\setlength{\parindent}{0pt}
\setlength{\parskip}{6pt plus 2pt minus 1pt}
}
\ifx\paragraph\undefined\else
\let\oldparagraph\paragraph
\renewcommand{\paragraph}[1]{\oldparagraph{#1}\mbox{}}
\fi
\ifx\subparagraph\undefined\else
\let\oldsubparagraph\subparagraph
\renewcommand{\subparagraph}[1]{\oldsubparagraph{#1}\mbox{}}
\fi

\let\rmarkdownfootnote\footnote%
\def\footnote{\protect\rmarkdownfootnote}

\usepackage{titling}

  \title{}
  \pretitle{\vspace{\droptitle}}
  \posttitle{}
  \author{}
  \preauthor{}\postauthor{}
  \date{}
  \predate{}\postdate{}

\usepackage{amsmath}
\usepackage{amsfonts,amssymb,textcomp,gensymb}
\usepackage{bbm}

\usepackage{scrpage2}
\ifoot[]{}
\ofoot[\pagemark]{\pagemark}

\usepackage{lscape}
\newcommand{\landscapebegin}{\begin{landscape}}
\newcommand{\landscapeend}{\end{landscape}}

\begin{document}

\section{Genetic control of plasticity of oil yield for combined abiotic
stresses using a joint approach of crop modeling and genome-wide
association}\label{genetic-control-of-plasticity-of-oil-yield-for-combined-abiotic-stresses-using-a-joint-approach-of-crop-modeling-and-genome-wide-association}

Brigitte Mangin\textsuperscript{1a}, Pierre
Casadebaig\textsuperscript{2a}, Eléna Cadic\textsuperscript{1}, Nicolas
Blanchet\textsuperscript{1}, Marie-Claude Boniface\textsuperscript{1},
Sébastien Carrère\textsuperscript{1}, Jérôme Gouzy\textsuperscript{1},
Ludovic Legrand\textsuperscript{1}, Baptiste
Mayjonade\textsuperscript{1}, Nicolas Pouilly\textsuperscript{1},
Thierry André\textsuperscript{3}, Marie Coque\textsuperscript{4}, Joël
Piquemal\textsuperscript{5}, Marion Laporte\textsuperscript{6}, Patrick
Vincourt\textsuperscript{1}, Stéphane Muños\textsuperscript{1}, Nicolas
B. Langlade\textsuperscript{1}

\textsuperscript{1} LIPM, Université de Toulouse, INRA, CNRS,
Castanet-Tolosan, France\\
\textsuperscript{2} AGIR, Université de Toulouse, INRA, INPT, INP-EI
PURPAN, Castanet-Tolosan, France\\
\textsuperscript{3} Soltis, Domaine de Sandreau, Mondonville F-31700
Blagnac, France\\
\textsuperscript{4} BIOGEMMA SAS, Domaine de Sandreau, Mondonville
F-31700 Blagnac, France\\
\textsuperscript{5} SYNGENTA SEEDS, 12 chemin Hobit, F-31790 Saint
Sauveur, France\\
\textsuperscript{6} RAGT 2n, Site de Bourran, F-12000 Rodez, France\\
\textsuperscript{a}These authors contributed equally to this work

\subsection[Abstract]{\texorpdfstring{Abstract\footnote{This is the
  post-print version of the manuscript published in \emph{Plant, Cell,
  and Environment}
  (\href{http://dx.doi.org/10.1111/pce.12961}{10.1111/pce.12961})}}{Abstract}}\label{abstract}

Understanding the genetic basis of phenotypic plasticity is crucial for
predicting and managing climate change effects on wild plants and crops.
Here, we combined crop modeling and quantitative genetics to study the
genetic control of oil yield plasticity for multiple abiotic stresses in
sunflower.

First we developed stress indicators to characterize 14 environments for
three abiotic stresses (cold, drought and nitrogen) using the SUNFLO
crop model and phenotypic variations of three commercial varieties. The
computed plant stress indicators better explain yield variation than
descriptors at the climatic or crop levels. In those environments, we
observed oil yield of 317 sunflower hybrids and regressed it with three
selected stress indicators. The slopes of cold stress norm reaction were
used as plasticity phenotypes in the following genome-wide association
study.

Among the 65,534 tested SNP, we identified nine QTL controlling oil
yield plasticity to cold stress. Associated SNP are localized in genes
previously shown to be involved in cold stress responses: oligopeptide
transporters, LTP, cystatin, alternative oxidase, or root development.
This novel approach opens new perspectives to identify genomic regions
involved in genotype-by-environment interaction of a complex traits to
multiple stresses in realistic natural or agronomical conditions.

\newpage

\subsection{Introduction}\label{introduction}

\emph{Adaptation to climate change requires new crop varieties adapted
to new management options.} Adaptation of agriculture is a key factor to
lessen the impact of climate change (Lobell et al., 2008). The crop
exposition to unfavorable growing periods can be partially controlled by
adapting crop management, i.e by shifting planting dates or choosing a
cultivar with an adequate phenology (Acosta-Gallegos and White, 1995).
But such adaptations also have side-effects and while flowering can be
successfully desynchronized from the period of occurrence of water
deficit, crop emergence would be more exposed to cold stress. In this
case, adaptation to climate change should also include the development
of new crop varieties (Rosenzweig et al., 1994), with new or improved
properties such as tolerance to cold or other abiotic stresses.
Moreover, because the multiplicity of cultivation conditions (soils,
climatic uncertainty), a single genotype can be exposed to random
unfavorable growing conditions within its cultivation area, ultimately
impacting the expected crop performance. To ensure a stable performance
under uncertain conditions, newly developed genotypes not only need to
be tolerant, i.e.~adapted to a single type of environment
(specialisation) but also plastic, i.e.~to be able to adapt in most of
growing conditions encountered in the targeted cultivation area
(Sambatti and Caylor, 2007).

\emph{Phenotypic plasticity is a key process for crop productivity under
climate change.} One way plants will respond to these changes is through
environmentally induced shifts in phenotype (phenotypic plasticity)
(Nicotra et al., 2010). While the process of phenotypic plasticity is
mostly studied on natural systems, its implications on crop productivity
under climate change are interesting for plant breeding (DeWitt and
Langerhans, 2004; Sadras et al., 2009). Empirically relationships
between plant traits and environmental variables (norms of reaction) are
known to vary within species in nature as well as in crop species, are
considered heritable traits themselves subject to natural or artificial
selection (Sambatti and Caylor, 2007; Via and Lande, 1985). However,
crop growth in a fluctuating environment generates complex and dynamic
interactions between plant and environment, under the control of
cultural practices.

\emph{It is necessary to unravel and measure abiotic stress levels
before assessing plasticity in plant traits.} In these conditions,
assessing plasticity in plant traits is limited by our capacity to
unravel those interactions and estimate abiotic stress levels at the
plant scale. Actually, each growth condition creates a unique
combination of those stress levels, with possible identical combinations
in different growth conditions. For example, crops growing in
continental climates might be exposed to both cold (during emergence)
and heat (during flowering) stresses; with high temperatures driving a
strong evaporative demand and water deficit, which also limits plant
nitrogen uptake from the transpired water stream (Kiani et al., 2016).
Accordingly, in a given cultivation area, stresses are not independent
(Vile et al., 2012) and need to be characterized and modelled prior
studying their impact on plant traits.

\emph{Crop simulation and modeling can help to characterize environment
from the crop point of view.} Because the environment is the largest
component of the phenotypic variability of most plant traits and of
course crop yield, its quantitative characterization is of major
importance (Lake et al., 2016). Crop simulation models are based on
mathematical equations representing the crop growth and development as a
function of environment (climate, soil and management). Such tools can
give access to plant-level state variables, such as time-series of
several abiotic stresses, in large range of growing conditions otherwise
difficult to characterize with sensors. This methodology was recently
implemented and allowed to identify major types of water deficit
patterns for rainfed wheat in the Australian target population of
environments (Chenu et al., 2013) or for coupled thermal and water
stress patterns for chickpea in Australian National Variety Trials (Lake
et al., 2016). In sunflower, a crop model was developed by Casadebaig et
al. (2011) and takes into account water, cold, heat, and nitrogen
stresses to estimate their impact on grain yield and oil content.

\emph{Genetic studies of genotype-by-environment interaction} (GxE)
Although plasticity has long been recognized as an interesting trait in
ecological and crop science since the pioneering work of Bradshaw in the
1950s, the identification of genetic variation involved in plasticity is
scarce and recent. An approach to deal with phenotypic variation due to
environmental effects is to develop multi-environmental QTL analysis to
identify QTLxE interactions as reviewed in Van Eeuwijk et al. (2010).
Although this approach was widely used for single or multiple traits,
this can provide a greater sensitivity for environment-dependent QTL,
inform on the stability of the QTL effect, but not on the environmental
factor(s) the QTL is responding to. To overcome this issue, several
studies compared two conditions, varying a single environmental factor
and were able to study the genetic control of the plasticity to a
specific stress for a particular trait (El-Soda et al., 2015; McKay et
al., 2008). Several genes involved in GxE have been cloned in plants
(reviewed by Des Marais et al., 2013) by using a combination of fine QTL
mapping and candidate gene approaches. Most genes are involved in
flowering time control (FT, PPD1, FLC, FRI, PHYC, CO). Examples
concerning abiotic stress responses are still rare: CBF2 for cold stress
(Alonso-Blanco et al., 2005), P5CS1 for osmotic stress (Kesari et al.,
2012), SUB1A for submersion in rice (Fukao et al., 2011) and RAS1 for
saline stress tolerance (Ren et al., 2010). No gene showing a GxE
interaction for drought stress nor for any yield-related trait were
cloned or fine-mapped yet. This certainly lies respectively in the
difficulty to characterize drought stress in natural conditions and in
the complex genetic architecture of yield-related traits, which implies
small effect mutations. Indeed successful examples were obtained with
easy-to-phenotype traits (flowering time) and easy-to-setup
environmental factors controlled one at a time.

This reductionist vision of the environment prove to be efficient for
gene cloning. However, more complex and realistic approaches are needed
to understand plant and crop responses to environmental factors and
therefore breed for trait plasticity ultimately providing stable crops.
In the context of the phenomics era (Großkinsky et al., 2015), with a
greater description of the environment, and with the genomic tools
accessible on many species, these approaches should flourish. However,
to our knowledge, no study tackled yet the identification of the genetic
control of plant plasticity to combined environmental factors in nature
or field trials (Mahalingam, 2015). Thanks to prior crop modeling, this
approach is now amenable, and shall take advantage of a precise
description of the plant stress factors, of the statistical power of
multi-environment trials and of the realistic nature of the measured
traits.

In this study, we developed a novel approach that combined crop modeling
and quantitative genetics to identify the genetic basis of oil yield
plasticity in sunflower. Our methodology consisted in estimating four
abiotic stress indicators in a range of environments using a crop model
and in selecting the indicators explaining the best the grain yield
plasticity of commercial varieties. The plant-level stress estimations
on these varieties were used to characterize each experimental sites for
the three selected abiotic stresses. We could then estimate the
plasticity of oil yield for each stress and for every hybrid of a
sunflower diversity panel cultivated in every site. Therefore, as a
demonstration on cold stress, we could successfully perform a
genome-wide association study to identify genomic regions putatively
involved in oil yield plasticity to cold.

\subsection{Material and methods}\label{material-and-methods}

\subsubsection{Plant material}\label{plant-material}

Association mapping was carried out on a panel of 317 inbred lines from
INRA and sunflower breeding companies. This panel was a subset of the
core collection of 384 inbred lines of Cadic et al. (2013) chosen for
its diversity from an initial set of 752 inbred lines (Coque et al.,
2008). It was comprised of both elite lines, parents of commercial
hybrids, and lines with introgressions from several wild Helianthus
accessions, including H. \emph{annuus}, H. \emph{argophyllus} and H.
\emph{petiolaris}.

Oil yield was observed on testcross progeny obtained by crossing panel
lines with testers according to their status (maintainers of cytoplasmic
male sterility {[}B-lines{]} or fertility restorers {[}R-lines{]}), as
described in Cadic et al. (2013). R-lines were crossed with the two CMS
PEFS71501 counterparts of B-line proprietary testers (FS71501 or AT0521)
while the B-lines were crossed with two R-line testers: 83HR4gms and
SOLR001M. 83HR4gms is derived from the 83HR4 line and was converted to
female by the introduction of a genetic male sterility, SOLR001M is a
proprietary line carrying PEF1 cytoplasmic male sterility (Crouzillat et
al., 1991; Serieys, 1984) which it maintains, although it is a restorer
for classical PEFS71501 cytoplasm (Table 1).

\subsubsection{Description of the multi-environment trial
(MET)}\label{description-of-the-multi-environment-trial-met}

From 2008 to 2010, eight locations located in the center and Southwest
of France were planted with the testcross progeny. In six location,
trials were conducted with and without irrigation, providing a total of
17 location x treatment x year combinations, (designated as
environments). The panel lines were evaluated on the same tester in each
environment (Table 1). Each experiment was an Augmented-Design (Federer,
1961) formed of blocks, with 24 or 30 entries replicated in two
sub-blocks. Each sub-block was randomized separately and contained two
to four control hybrids.

\begin{quote}
\textbf{Table 1. Details on location, treatment, year, testers and
observed hybrids for the 17 environments.}
\end{quote}

The climatic variability on experimental locations was summarized by
computing the mean air temperature, the sum of water inputs (rainfall
and irrigation) and the climatic water deficit (difference between
precipitations, P and potential evapotranspiration, PET) on the cropping
period (see supplementary Figure S1 and Table S1). No environment had a
climatic water deficit (ranging from -177 to -458 mm), meaning that the
climatic evaporative demand was always above the water supply (even
accounting for irrigation). Rainfall on the cropping period ranged from
a low 36 mm to 368.5 mm and the average amount of irrigation was 74 mm.
Trials were performed on various soils depth leading to a soil water
capacity (SWC) from 112 mm to 240 mm. Mean nitrogen fertilization was 60
kg/ha (eq. mineral nitrogen).

Among the 17 environments, we discarded 3 environments (AI09\_I,
AI09\_NI, CO09\_NI). The first two were outliers for the observed yield
and the SUNFLO model failed at simulating yield phenotypes close to the
observed one for the controls. The last one did not exhibit genotypic
effect for the panel phenotypes, the estimated genotypic variance was
judged non-significantly different from zero by a Z-ratio test in the
naïve linear mixed model used to correct for micro-environment effects
and to predict the BLUP genotypic value of the yield panel lines.

\subsubsection{Intra environment phenotypic data
analysis}\label{intra-environment-phenotypic-data-analysis}

Within each environment, the oil yield was adjusted for
micro-environment effects using ASReml-R (Butler et al., 2009), as
described in Cadic et al. (2013). A linear mixed model (named naïve in
Cadic et al., 2012) with a random effect for the genotypic value of the
panel lines, including blocks and sub-blocks as fixed effects was
compared to two spatial models. The spatial models included (a) random
effects of row and column or (b) a first-order autoregressive process in
the residuals to take into account autocorrelation between neighbour
plots. The three models were compared using the Akaike criterion (AIC)
and the best linear unbiased predictors (BLUP) of the genotypic values
were extracted from the best model according to AIC, for the next step
of analysis.

\subsubsection{Estimation and choice of abiotic stresses in each
environment using
simulation}\label{estimation-and-choice-of-abiotic-stresses-in-each-environment-using-simulation}

SUNFLO is a process-based model for the sunflower crop which was
developed to simulate the grain yield and oil concentration as a
function of time, environment (soil and climate), management practice
and genetic diversity (Casadebaig et al., 2011; Debaeke et al., 2010;
Lecoeur et al., 2011). The model simulates the main soil and plant
processes: root growth, soil water and nitrogen content, plant
transpiration and nitrogen uptake, leaf expansion and senescence and
biomass accumulation, as a function of main environmental constraints
(temperature, radiation, water and nitrogen deficit).

This model is based on a conceptual framework initially proposed by
Monteith (1977) and now shared by a large family of crop models (Brisson
et al., 2003; e.g. Jones et al., 2003; Keating et al., 2003). In this
framework, the daily crop dry biomass (DM\(_t\)) is calculated as an
ordinary difference equation (Equation 1) function of incident
photosynthetically active radiation (PAR, MJ m\textsuperscript{-2}),
light interception efficiency (\(1-\exp^{-k \textrm{ LAI}_t}\)) and
radiation use efficiency (RUE\(_t\), g MJ\textsuperscript{-1}, Monteith
(1994)). The light interception efficiency is based on Beer-Lambert's
law as a function of leaf area index (LAI\(_t\)) and light extinction
coefficient (\(k\)).

\begin{equation}
  \textrm{DM}_t = \textrm{DM}_{t-1} + \textrm{RUE}_t ~ (1-\exp^{-k \textrm{ LAI}_t}) \textrm{ PAR}_t
\label{eq:DM}
\end{equation}

Broad scale processes of this framework, the dynamics of LAI,
photosynthesis (RUE) and biomass allocation to grains were split into
finer processes (e.g plant phenologic development, leaf expansion and
senescence, response functions to environmental stresses) to reveal
genotypic specificity and to allow the emergence of GxE interactions.
Globally, the SUNFLO crop model has about 50 equations and 64 parameters
(43 plant-related traits and 21 environment-related). When evaluated on
the presented MET dataset, the SUNFLO model predicted accurately the
performance of control hybrids across environments: the root of mean
square error (RMSE) was 0.3 t ha \textsuperscript{-1}, relative RMSE was
9.6 \%, bias was -0.14 t ha\textsuperscript{-1}.

Using the SUNFLO model, we computed two indicators (continuous and
discrete) per type of considered abiotic stresses to characterize the
different environments (Table 2). Each indicator was integrated over
three key periods: vegetative stage (veg), flowering period (flo) and
grain filling period (fil). We also considered the sum over two periods
and during the whole cropping period, for a total of seven time periods
per indicator. All these indicators corresponded to the mean stress felt
by the control hybrids during the above periods.

For water stress, the fraction of transpirable soil water (FTSW)
represents yield limitation through water deficit (integration of 1
minus FTSW); ETR is the conditional sum of days, if the ratio of the
real evapotranspiration (ET) to potential evapotranspiration (PET) was
less than 0.6 (threshold for photosynthesis limitation). For cold
stress, LTs is the conditional sum of days if mean air temperature was
below 20°C and LTi represent low temperatures impact on photosynthesis
(integration of 1 minus equation 2). Heat stress indicators were
computed following the same logic, albeit representing high temperatures
impact on photosynthesis. Equations (2) and (3) are used in the crop
model to define the radiation use efficiency (RUE) response to
temperature (Villalobos et al., 1996). For nitrogen deficit, NAB is the
amount of absorbed nitrogen in the considered cropping period and NNI is
the sum of 1 minus nitrogen nutrition index, which indicates crop
nitrogen deficit (Debaeke et al., 2012; Lemaire and Meynard, 1997).

\begin{equation}
\textrm{LTRUE} = \left\{
  \begin{array}{ll}
  0 & \textrm{if $T_m < T_{b}$} \\
  \frac{T_m}{T_{ol} - T_b} - \frac{T_b}{T_{ol} - T_b} & \textrm{if $T_b < T_m < T_{ol}$} \\
  1 & \textrm{else}
  \end{array}
  \right.
  \label{eq:LTRUE}
\end{equation}

\begin{equation}
\textrm{HTRUE} = \left\{
  \begin{array}{ll}
  1 & \textrm{if $T_m < T_{ou}$} \\
  \frac{T_m}{T_{ou} - T_c} - \frac{T_c}{T_{ou} - T_c} & \textrm{if $T_{ou} < T_m < T_c$} \\
  0 & \textrm{else}
  \end{array}
  \right.
  \label{eq:HTRUE}
\end{equation}

where \(T_m\) denotes the mean daily air temperature (°C), \(T_b = 4.8\)
is the base temperature (°C), \(T_{ol} = 20\) is the optimal lower
temperature (°C), \(T_{ou} = 28\) is the optimal upper temperature (°C)
and \(T_c = 37\) is the critical temperature (°C).

\begin{quote}
\textbf{Table 2. Description of abiotic stress indicators simulated by
the crop model.}
\end{quote}

Three sunflower varieties (Melody, Pacific, Pegasol) which were used as
controls within the environments had their phenotypic characteristics
previously included in the SUNFLO model (Supplementary Table S2). The
soil characteristics, the crop management and climate data were
collected to allow simulation for each environment. One soil
characteristic, the soil depth, is difficult to observe and has a strong
impact on the simulated yield. Instead of using the approximated value
given by the breeders and the farmers, we adjusted this parameter by
minimizing the empirical mean square error between the observed and the
simulated yields.

Model choice to select the stress indicators was made with the AIC using
the native R function
\href{https://stat.ethz.ch/R-manual/R-devel/library/stats/html/lm.html}{lm}.
AIC of each model was computed for all three control varieties and model
choice was made on the mean over the three controls. Compared linear
models, which fitted the grain yield with the indicators, were limited
to combinations of only one indicator per type of stresses. All models
integrating one, two, three or four stresses were compared; in total
50,624 models were computed and compared.

The R function lm was also used to compute the p-value of the Fisher
test for each indicator including in the best model.

\subsubsection{Genetic study}\label{genetic-study}

\paragraph{Estimation of plasticities of oil yield to water, nitrogen
and cold stresses in the diversity
panel}\label{estimation-of-plasticities-of-oil-yield-to-water-nitrogen-and-cold-stresses-in-the-diversity-panel}

In order to get plasticity phenotypes that reflect the responses of the
panel lines to the different abiotic factors, we adjusted the BLUP
phenotype of each panel line with the following linear model:

\begin{equation}
  Y_{ij} = a_i + b_i ~ DS_j + c_i ~ CS_j + d_i ~ NS_j + \epsilon_{ij}
\label{eq:blup}
\end{equation}

where \(Y_{ij}\) is the BLUP of the phenotype for the \(i\)th genotype,
\(a_i\) the potential phenotype in an environment with average stresses,
\(DS_j\) the water stress indicator in the \(j\)th environment
calculated as the mean over the 3 controls, \(b_i\) the slope linked to
the water stress, \(CS_j\) the cold stress indicator in the \(j\)th
environment calculated as the mean over the 3 controls, \(c_i\) the
slope linked to the cold stress, \(NS_j\) the nitrogen stress indicator
in the \(j\)th environment calculated as the mean over the 3 controls,
\(d_i\) the slope linked to the nitrogen stress and \(\epsilon_{ij}\)
the residual variance. The covariates \(DS\), \(CS\), and \(NS\) were
centered before computing the regression model. The three estimated
covariate coefficients \(\hat{b}_i\), \(\hat{c}_i\), and \(\hat{d}_i\)
of this model are the plasticity phenotypes of interest, i.e.~the
genotype slopes in response to water, cold and nitrogen stresses
respectively. To compare model (\ref{eq:blup}) to a more simple
regression model, we also fitted the BLUP phenotype of each panel line
using the cold stress indicator as a single regressor.

Before computing the above linear model, the oil yield missing data were
imputed using the
\href{http://cran.r-project.org/web/packages/missMDA/index.html}{missMDA}
R package (Josse et al., 2012; Josse and Husson, 2016). All recorded
traits (24 traits for a total of 198 traits x environments) were used to
impute missing oil yield values. However, panel lines with less than six
observations per oil yield trait over the environments were discarded.

\paragraph{Combined stresses in a single multi-stress index of
plasticity}\label{combined-stresses-in-a-single-multi-stress-index-of-plasticity}

In order to compare the panel lines for the stability against multiple
stresses together, we defined a multi-stress plasticity index accounting
for the variance-covariance of the stress slopes, equals to:

\begin{equation}
  (\hat{b}_i , \hat{c}_i , \hat{d}_i ) V^{-1} (\hat{b}_i , \hat{c}_i , \hat{d}_i)^t
\label{eq:multistress}
\end{equation}

where \(V\) denotes the (3 by 3) variance-covariance matrix.

To visualize the strategy of the panel lines versus the three stresses,
we drew a star representation using the
\href{https://stat.ethz.ch/R-manual/R-devel/library/graphics/html/stars.html}{star}
function of the R package graphics. The three plasticity phenotypes were
first taken in absolute value and scaled to 0-1 in order to have three
comparable values with values close to 0 noted the stability and values
close to 1 noted the instability. Then they were normalized by panel
line to sum to 1 in order to get comparable values for all panel lines.
The star representation was applied on these scaled and normalized
values that can be interpreted as the percentage of stability dedicated
to each stress.

\paragraph{Genotyping and map
building.}\label{genotyping-and-map-building.}

A set of 197,914 SNP were used to produce an AXIOM® genotyping 96-array
(Affymetrix, Santa Clara, CA, USA). These SNP were selected from either
genomic re-sequencing or transcriptomic experiments. An additional set
of 6,800 non-polymorphic sequences were added as controls. Combined with
internal technical controls, the AXIOM® genotyping 96-array was designed
with a total of 445,876 probesets.

Genomic DNA from the 317 panel lines and two recombinant inbred lines
(RIL) populations INEDI and FUxPAZ2 obtained from the cross between XRQ
and PSC8 lines (180 RIL) and from the cross between FU and PAZ2 lines
(87 RIL) respectively, were genotyped with the AXIOM® array. All
hybridization experiments were performed by Affymetrix and the genotypic
data were obtained with the GTC software (Affymetrix). From the 197,914
SNP, 35,562 were polymorphic between XRQ and PSC8 and 28,529 between FU
and PAZ2.

We used CarthaGène v1.3 (Givry et al., 2005) to build the genetic maps
of the INEDI population and the FUxPAZ2 population separately. For the
INEDI population, we added to the set of AXIOM® SNP the markers
previously mapped by Cadic et al. (2013) which allowed completing this
map and assigning AXIOM® markers to appropriate linkage group (LG). We
built a consensus map with common markers of both previous maps using
Biomercator v4.0 (Sosnowski et al., 2012) and we projected the specific
markers of the two previous maps on this consensus map. Unmapped AXIOM®
SNP were placed by BLAST analysis on the RHA280xRHA801 genetic map based
on 454 sequencing (Kane et al., 2011) and finally projected on the INEDI
and FUxPAZ2 consensus map. The remaining AXIOM® SNP were located by
computing the linkage disequilibrium (LD) measurements proposed by
Mangin et al. (2012). They were mapped to the same position of the
mapped SNP in maximum LD. For this, we used as LD statistics the maximum
of the \(r^2_V\) and \(r^2_{VS}\) measurements that correct for
relatedness and for both structure and relatedness, respectively.

\paragraph{Association tests}\label{association-tests}

Association mapping was based on a set of 65,534 SNP with MAF
\textgreater{} 0.05. Similarly to our previous work (Cadic et al.,
2013), two association models were performed using EMMA (Kang et al.,
2008) based on the Yu et al. (2006) model. Both models included a
correction for the genomic relatedness using the alike-in-state (AIS)
kinship estimated with EMMA version v1.1.2 R package (Villanova et al.,
2011) using all the above SNP. The population structure modelled as the
restorer or maintainer status of the panel lines was added in the second
model leading to the following model:

\begin{equation}
  Sl_i = \sum_c X_{ic} ~ \alpha_c + M_{il} ~ \theta_l + u_i + e_i
\label{eq:structure}
\end{equation}

\(Sl_i\) is the slope for \emph{i}th line, \(X_{ic}\) is the line
status, \(\alpha_c\) is the effect of the line status \(c\), \(M_{il}\)
is the genotype of the \emph{i}th line at locus \emph{l}, \(\theta_l\)
is the effect of locus \emph{l}. \(\alpha_c\) and \(\theta_l\) are
considered to be fixed effects. \(u_i\) is the random polygenic effect
modeling genetic relatedness with
\(\textrm{Var}(u) = \sigma^2_u ~ K_{ais}\) where \(K_{ais}\) is an AIS
matrix and \(\textrm{Var}(e) = \sigma^2_e ~I\) where \(I\) denotes the
identity matrix. Multiple testing correction was achieved using an
approximate effective number of tests (\(M_{eff}\)) based on the
eigen-values of the SNP correlation matrix as proposed by Li and Ji
(2005). We computed \(M_{eff}\) by using blocks of 250 markers and
assuming that the blocks were independent.

In order to conduct a multi-loci analysis, the genotypic data was
imputed using Beagle v4.0 (Browning and Browning, 2007) and MLMM (Segura
et al., 2012) was used for this purpose. We stopped the forward approach
of MLMM when the variance of the polygenic term was non significant,
which is a little different compared to the initial forward procedure of
MLMM that stops when the estimator of this polygenic variance is equal
to 0. The significance of the polygenic variance component was judged
with a log-likelihood ratio test and a risk of 1\%. This log-likelihood
ratio test compared the model with and without the polygenic effect. Two
times the difference between the log-likelihood of the models is known
to follow asymptotically a mixture between a Dirac at 0 and a Chi-square
distribution with 1 degree of freedom (Self and Liang, 1987). When the
forward approach stopped, a SNP was judged associated if its Bonferonni
corrected p-value, using \(M_{eff}\) as the number of independent tests,
was inferior to the chosen type I error. The associated SNP were put all
together with the polygenic term to create the final multi-loci linear
model. The associated SNP effect, their reevaluated p-values were
computed with the base R function lm in this final model (R Core Team,
2014).

\subsubsection{Functional annotation of associated
SNP}\label{functional-annotation-of-associated-snp}

Context sequences of associated SNP were compared to the sunflower
genome (line XRQ) sequenced using the PacBio (Pacific Biosciences, Menlo
Park, CA, USA) technology
(\url{https://www.heliagene.org/HanXRQ-SUNRISE/}) by blast analysis. In
case of ambiguous positioning on the genome, we retained the chromosomal
position in accordance to genetic map location of the marker. Similarly,
we positioned associated SNP on the reference transcriptome
(\url{https://www.heliagene.org/HaT13l/}) All SNP were situated in gene
coding regions, the sunflower genes, the best Arabidopsis hit with its
description in TAIR10 are indicated in Table 5.

\subsection{Results}\label{results}

\subsubsection{Estimation of abiotic stresses in field
environment.}\label{estimation-of-abiotic-stresses-in-field-environment.}

The characterization of abiotic stresses at the plant level through
modeling and simulation explained observed yield varibility better than
climate-based indicators (Figure 1). For example, the correlation
between yield observed in field experiments and water deficit computed
from (1) climate data only (P - PET), (2) climate, soil, and management
data (P - PET + SWC + irrigation) and (3) simulated plant data (ETR,
defined in Table 2) was gradually stronger (respectively -0.01, -0.65
and -0.86). The correlation between yield and water deficit index
increased in strength with the ``proximity'' of the regressor to the
plant, revealing the expected negative impact of water deficit on crop
yield.

\begin{quote}
\textbf{Figure 1. Relation between observed grain yield and several
abiotic stress indicators}
\end{quote}

The correlation between different type of abiotic stresses (Figure S2),
indicated that cold (LT) and heat stresses (HT) were naturally the most
highly negatively correlated combination. The nitrogen indicator NAB was
not correlated to other indicators except the negative correlation with
the water stress indicator (ETR, particularly during grain filling). The
other nitrogen indicator NNI, which was not correlated to NAB showed
both a positive correlation with heat stress indicators and a negative
correlation with cold indicators. The water stress indicator FTSW during
grain filling was also positively correlated to heat stress indicators
(HT), nitrogen stress (NNI), and negatively correlated to cold stress
indicators. Within a single type of stress, correlations between
indicators computed during different cropping periods were generally
positive and high (Figure S2), ranging 0.40 for water deficit (FTSW) to
0.98 for heat stress (HTi), with nitrogen and cold indicators
in-between.

\textbf{Figure 2. Heat map of the correlations between stress
indicators.}

\subsubsection{Model selection to estimate the best combination of
abiotic
stresses.}\label{model-selection-to-estimate-the-best-combination-of-abiotic-stresses.}

The model selected as having the best AIC on average over the control
genotypes was a three indicator model, including a cold stress indicator
during the vegetative stage (LTi\(_{\textrm{veg}}\)), a water stress
indicator during the vegetative and the flowering period
(FTSW\(_{\textrm{veg+flo}}\)), and a nitrogen indicator during the whole
growth period (NAB). Table 3 presents the p-value of the Fisher test for
each control in this best model and the proportion of variance explained
by this best model for each control.

\begin{quote}
\textbf{Table 3. Results of the regression between grain yield and
abiotic environmental indicators: water, cold and nitrogen stresses.}
\end{quote}

Each stress indicator had a significant effect on the grain yield for
each control genotype. The highest impact was due to cold stress for
Melody (p-value = 2.39 10\textsuperscript{-4}) and the smallest was due
to water stress for Pacific (p-value = 2.91 10\textsuperscript{-2}) as
indicated in Table 3. All together these indicators explained very well
the grain yield variability of each control since the percentages of
explained variance were 90\%, 93\% and 94\% for Pegasol, Pacific and
Melody respectively.

Although cold stress has a strong impact on yield (-0.35 q
h\textsuperscript{-1} d\textsuperscript{-1}), the relatively low number
of day of cold stress in the MET (8.7 on average, variation in the MET
shown on Figure 3) reduces its impact, whereas water stress impacts the
most yield due to the high number of days of stress (33.9 on average)
even if sunflower is relatively tolerant (-0.14 q h\textsuperscript{-1}
d\textsuperscript{-1}).

\begin{quote}
\textbf{Figure 3. Ranges of variation of abiotic stresses in the
multi-environment trial.}
\end{quote}

\subsubsection{Plasticity in the diversity
panel.}\label{plasticity-in-the-diversity-panel.}

As we described the actual cold, nitrogen and water stresses felt by
sunflower in 14 environments, we also measured oil yield in a diversity
panel in those environments. This allowed us to compute plasticities of
oil yield to abiotic stresses for each line in the panel as the slopes
of regression of oil yield to individual stress indicators. These three
slopes represented how plastic was the response of a line faced to a
given stress and are therefore referred as plasticity phenotypes later.
Minimum, mean, maximum, and variance of these plasticities as well as
their correlation are presented in Table 4 and their distribution
histograms are in supplementary Figure S2. Cold stress plasticity
appears genetically independent to water and nitrogen stress
plasticities. On the contrary, nitrogen and water stress plasticity are
highly correlated (Pearson correlation of 0.62) suggesting common
genetic control of these traits in our panel.

\begin{quote}
\textbf{Table 4. Variation and correlation of the plasticities of oil
yield for different abiotic stresses in the diversity panel.}
\end{quote}

To illustrate the importance of multi-stress indicator modelling, we
compared regression of oil yield in a single stress indicator model
(including only cold stress indicator) and in the multi-stress indicator
model in Figure 4. Oil yields of the most sensitive and tolerant lines
were plotted against the cold stress indicator. Point clouds are closer
to the regression lines in the multi-stress indicator model indicating a
better characterization of cold stress impact in this modelling
approach.

\begin{quote}
\textbf{Figure 4. Regression of oil yield against the cold stress
indicator for the most tolerant (green triangles) and sensitive (red
squares) panel lines showing the ability of multi-stress modelling to
better characterize the environment.}
\end{quote}

\paragraph{A multi-stress plasticity
index.}\label{a-multi-stress-plasticity-index.}

Following the estimation of the three single stress plasticities, we
were interested in calculating a multi-stress plasticity index to
describe the general abiotic tolerance of every line. This index is a
weighted sum of the three abiotic stress plasticities taking into
account the correlation between them, specifically the one between
drought and nitrogen. This index allowed allowed us to rank the panel
lines and to describe the different strategies observed in the panel to
tolerate combined stresses. This is illustrated on Figure 5 that shows
the multi-stress plasticity index of every panel lines against its mean
oil yield in the MET. Stable panel lines, with a small multi-stress
plasticity index, as well as panel lines belonging to the border of the
point cloud were highlighted by a star representation (a triangle in our
case) illustrating their abiotic stress tolerance strategy. First, we
observed that unstable panel lines were generally more sensitive to cold
stress compared to stable lines. This was confirmed by a significant
correlation between the multi-stress plasticity index and the percentage
of plasticity due to the cold stress (p-value of 6
10\textsuperscript{-4}) although this was not observed for nitrogen or
water stress (p-value of 0.28 and 0.10 respectively). Second, we could
not identify a consensus strategy for stable panel lines. As examples:
(i) the three similarly stable panel lines (multi-stress index around
45) having high oil yield (around 15) developed tolerance to all
stresses but showed different plasticity patterns, (ii) the two most
stable panel lines had opposite profile of stability with the most
stable being sensitive to cold and the second to water.

\begin{quote}
\textbf{Figure 5 Relation between the multi-stress plasticity index and
the average oil yield showing differences in stress tolerance strategies
in the diversity panel of sunflower lines.}
\end{quote}

\subsubsection{Genetic map.}\label{genetic-map.}

In order to position the genomic regions controlling the abiotic stress
plasticity, we constructed a genetic map with the markers genotyped on
the diversity panel using three RIL populations. The final genetic map
was composed of 89,979 markers positioned in 4,782 genetic positions for
a total distance of 1,398.5 cM. On this map, 4,094 markers were used to
build the consensus map between the INEDI (XRQxPSC8) and FUxPAZ2 RIL
populations. Among the other markers, 27,663 were mapped using the INEDI
population alone, 13,807 were mapped using the FUxPAZ2 population alone,
29,586 were located thanks to a genotyping by sequencing map on the RIL
population RHA801xRHA280 (Kane et al., 2011) and the remaining 14,829
markers were placed by linkage disequilibrium. Details of these maps can
be found in the supplementary Table S3.

\subsubsection{Association study of oil yield plasticity to cold
stress.}\label{association-study-of-oil-yield-plasticity-to-cold-stress.}

For the genetic analysis, we focused, as a proof of principle of the
approach, on the plasticity to cold stress as it is the most impacting
stress and appeared genetically independent from the others. Among the
65,534 association tests, the effective number of tests (Li and Ji,
2005) was estimated around 14,000. Using this effective number of tests,
we kept SNP associated with a Bonferroni corrected p-value of 7.1
10\textsuperscript{-6} for a family wise type I error of 10\%. Only 2
SNP were detected by a marker by marker association analysis using EMMA.
The most significant SNP was located at the end of LG 5 in a QTL named
LG05.64, and was detected with a model including the maintainer or
restorer status as a fixed structure effect. The second is located at
the center of LG 17 and was detected without the line status effect.

We completed this study by a forward approach of the multi-loci
association analysis (MLMM) for both models (with or without the
maintainer/restorer status). Both models stopped with 6 SNP among which
four were judged associated. None of them was common between both models
but one SNP corresponded to the previously found SNP in LG05.64. The
most associated SNP was located at the end of the LG13. In total 9 QTL
could be identified: two located on chromosomes 5 and 10 and one on
chromosomes 9, 13, 14, 16 and 17. Their phenotypic effects on cold
stress plasticity varies from 10 to 21\% of the average plasticity in
the panel (Table 5).

\begin{quote}
\textbf{Table 5. Positions and estimated effects of SNP and genes
associated with oil yield plasticity to cold stress.}
\end{quote}

\subsubsection{Genes located in QTL controlling oil yield plasticity to
cold
stress.}\label{genes-located-in-qtl-controlling-oil-yield-plasticity-to-cold-stress.}

We were interested in genes containing associated SNP to link the
genetic identification to molecular and physiological processes
putatively involved in cold tolerance. All associated SNP were located
within coding sequences as expected from the AXIOM® genotyping array
design. Functional annotation of the corresponding genes pointed out
homologues of: NPF3.1, LTP, CYS6, NPF5.3, GMII, RPD1, PPX1, HAOX2 and
IAR4 (from the most to the least significantly associated, as shown in
Table 5). Strikingly, two close homologues of oligopeptide transporters
(NPF3.1 and NPF5.3) are present in associated QTL on LG 5 and 14 and two
homologues of genes involved in root development (RPD1 and IAR4) on LG 9
and 17. In addition, homologues of a lipid transfer protein, a cystatin,
an alpha-mannosidase, a protein phosphatase and an aldolase are also in
QTL controlling oil yield plasticity to cold stress.

\subsection{Discussion}\label{discussion}

In this work, we developed a novel method to characterize the abiotic
stress levels on different environments by using crop modeling and
simulation and subsequently exploited it to identify genetic control of
stress plasticity. We implemented this environment characterisation
method on 17 locations from a multi-environment trial for sunflower and
four abiotic stresses (water, nitrogen, cold and heat). The SUNFLO model
(Casadebaig et al., 2011) was used to simulate stress patterns dynamics
for three varieties used as controls in each location and integrated
indicators were computed from these data considering different crop
phenological stages and physiological processes (eight stress indicators
over seven periods). We used a model selection approach to select the
best linear model among combinations of stress indicators used as
regressors for yield. Water, nitrogen and cold stresses were retained as
the most explicative abiotic stresses for yield variability in this MET.
Using reaction norms as conceptual reference, we computed abiotic stress
plasticities as the slopes of the linear regression of oil yield on
selected stress indicators. We then conducted an association study with
a panel of 317 lines genotyped for nearly 65,000 markers on oil yield
plasticity for cold stress as its was the most impacting stress (per
time unit) and was not correlated to other abiotic stress plasticities.

\subsubsection{Crop modelling helped to analyze abiotic stress patterns
and to explain their impact on
yield.}\label{crop-modelling-helped-to-analyze-abiotic-stress-patterns-and-to-explain-their-impact-on-yield.}

In a location, stress indicators can be climate-based (precipitations
minus evapotranspiration), crop-based (simple water balance including
soil water capacity and irrigation) or plant-based (simulated dynamic
water balance). We observed that the observed grain yield was best
explained by plant-based stress indicators because the interactions
between climate, leaf area dynamics, plant stomatal conductance
(isohydric vs anisohydric behaviours, e.g. Casadebaig et al. (2008) for
sunflower) and management practices can be partly reproduced by the crop
model algorithm. This is for example illustrated by environment
rankings, where some irrigated locations (CO08\_I, CO09\_I) can still
display a high level of water stress while a rainfed one (CA10\_NI,
VE09\_NI) show reduced water deficit.

Abiotic stresses also do not have the same impact on crop physiology
according to their timing of occurrence during the crop cycle (Table 3).
Among the seven possible combinations between main crop phases
(i.e.~vegetative, flowering, grain filling), we indeed observed that the
relevance of these timings was specific to the type of abiotic stress.
For cold stress, the detection of early crop growth (vegetative period)
was expected because this stress occurrence is strongly determined by
the climate (low temperature during crop installation). However, in
continental climates, where sunflower is mainly grown, we can also
observe cold temperatures at the end of the crop cycle. For water
stress, where interactions between crop growth and climate variability
are more important, vegetative and flowering periods were identified,
which is consistent with numerous previous reports on sunflower
(Blanchet et al., 1990; Cabelguenne et al., 1999). Regarding nitrogen
stress, the importance of this process over the whole crop cycle was
highlighted. Indeed, recent reports indicated that post-flowering
nitrogen absorption could also be significant (Andrianasolo et al.,
2016). Remarkably, heat stress was not identified as a major
contribution to yield variability in the MET. Actually, depreciative
effect of high temperatures on photosynthesis were caused by
temperatures that were almost never reached in our experimental
conditions. According to the current parameterization of the simulation
model, heat stress indicator (\ref{eq:HTRUE}) was only significant in
one location (CO09) and null (9/17 locations) or weak in the others
(Figures 2 and S1).

\subsubsection{Genetic control of plasticity of oil yield to cold
stress}\label{genetic-control-of-plasticity-of-oil-yield-to-cold-stress}

The most and fourth most associated SNP pointed to two genes highly
homologous to NPF3.1 and NPF5.3 that are both oligopeptide transporters.
These two independent association signals on chromosomes 5 and 14
strongly suggest a role of oligopeptide transport in tolerance to cold
stress observed in our experimental conditions i.e.~when young plants
are exposed to chilling. In plants, these transporters are key players
in nitrogen nutrition and therefore plantlet growth. The importance of
oligopeptide transport to tolerate cold is corroborated by the
demonstrated molecular adaptation of this transporter family in
antarctic icefish (\emph{Chionodraco hamatus}) adapted to sub-zero
temperatures (Maffia et al., 2003; Rizzello et al., 2013). The role in N
nutrition of these transporters in animal and plants indicates that
nutrient transport can be a limiting factor at low temperature that
likely limits remobilization of seed stocks and/or absorption and
transport of N from roots to aerial organs.

The second most associated SNP is located in a putative lipid transfer
protein (LTP) (QTL LG16.48). Many LTP have been reported to be
transcriptionally induced by freezing (reviewed in Liu et al., 2015) and
over-expression of LTP3 provided freezing tolerance in A.
\emph{thaliana} (Guo et al., 2013). This action could be due to membrane
stabilization as demonstrated by Hincha et al. (2001) in preventing
chloroplastic damages induced by freezing, or through its role in seed
lipid mobilization during germination and seedling growth (Pagnussat et
al., 2015) as shown in sunflower for another LTP (Pagnussat et al.,
2009). Another candidate genes is the cystatin CYS6 homologue located in
QTL LG13.72. Several homologues of cystatin were shown to be induced
during cold exposure in barley (Gaddour et al., 2001), maize (Massonneau
et al., 2005), wheat (Talanova et al., 2012), and increase, when
over-expressed, cold tolerance in Arabidopsis (Zhang et al., 2008).
Based on Prins et al. (2008), the sunflower cystatin could provide a
better regulation of Rubisco turnover in chloroplasts in cold
conditions. The alternative oxidase HAOX2 homologue found in QTL LG10.34
constitutes another good candidate gene for cold tolerance. The
Alternative Oxidase Pathway (AOP) has been described in many plants to
be involved in cold stress response as a biochemical protection against
overproduction of Reactive Oxygen Species due to the cold inhibition of
the electron transport chain in mitochondria (Feng et al., 2008).
Furthermore, the AOP was shown to participate in differential
cold-sensitivity between two maize genotypes (Ribas-Carbo et al., 2000).

Interestingly, Interestingly, two genes (homologues to RPD1, IAR4) were
found (QTL LG09.27, LG17.49) and their Arabidopsis counterparts share
similar features: both are involved in root development and both mutants
show temperature-sensitive phenotypes (at 20°C and 28°C) (Konishi and
Sugiyama, 2006; Quint et al., 2009). This suggests that root setting
could also be temperature- and genotype-dependent in sunflower. All
together, the functional annotation of QTL associated to cold stress
plasticity of oil yield identified several candidate genes and
physiological processes. Most of them were already described to be
involved in cold stress tolerance in other plants which supports our
study and indicates a probably short genetic distance between associated
SNP and causal mutations.

To complete our understanding of how these processes that act during the
early growth phase of sunflower, impact the final seed yield, further
studies putting in relation dynamic measurements of plant growth rate
and cell physiology would be enlightening.

\subsubsection{Phenotypic plasticity and tradeoff for potential
yield}\label{phenotypic-plasticity-and-tradeoff-for-potential-yield}

In the studied multi-environment network, mean oil yield was positively
correlated with a high sensitivity to environmental stresses, indicating
that a global gain in performance was generally associated to a higher
yield instability (Figure 5). This is globally in accordance with
previous claims that an increase in phenotypic plasticity allowed to
achieve better yield stability across environments but at the expense of
greater performance in low stress conditions (Sadras et al., 2009).
However the presence of both high-yielding and stress-tolerant genotypes
suggests this general observation can be genetically by-passed and
leaves room for more efficient sunflower varieties.

Stability of complex traits such as yield or fitness depends on
plasticity of numerous intermediate traits (likely physiological and
developmental processes) that are yet unknown. In our approach, we
studied directly the plasticity of the complex trait with the idea of
stabilizing it. The molecular and physiological processes pointed out by
the genetic analysis allow us to identify some candidate processes:
oligopeptide transport, root development, ROS scavenging, chloroplast
and mitochondrial physiology (i.e.~intermediate traits). In this kind of
approach, we can wonder whether we could detect key regulators such as
transcription factors (none in our case). Indeed, genetic variation in
those would likely have trade-off effects on various physiological
processes. Then, they would impact intermediate traits significantly but
with possibly opposite effects and at the end, no significant impact on
the resulting complex trait.

On the breeding strategy point of view, the considered environmental
stresses (temperature, nitrogen and water) do not necessarily coexist in
the French target population of environments: e.g.~south-western
production regions are exposed to drought and heat stress while the
temperatures in northern regions are low enough and necessitate short
crop cycle cultivars. This lack of spatial superposition of
environmental constraints allow to exploit the differential
sensitivities in the studied genetic material and adapt cultivar choice
or breeding according to local growing environment. On the other hand,
breeding for phenotypic plasticity of intermediate traits would
potentially result in resilience to increasingly unpredictable
environments (Nicotra et al., 2010).

\subsection{Conclusions}\label{conclusions}

Improving crop performance in low-input cropping systems requires a
coordinated improvement of genotypes and agronomical practices (Sadras
and Denison, 2016). In these growth-limiting conditions, abiotic
stresses occur in combined and dynamic patterns. Therefore,
disentangling those using models allows to understand their specific
impacts on complex traits (such as yield) and the genetic factors
potentially reducing those.

In our study, precise characterization of water, cold, heat and nitrogen
stresses allowed accurate identification of nine QTL and underlying
genes controlling stress plasticity. This joint approach between crop
modeling and quantitative genetics also permitted estimations of allelic
variation in natural conditions.

Such inter-disciplinary approach should be useful to conduct different
breeding strategies and adapt crop to climate variability through local
adaptation: maximizing performance in a given environment type, or
through global adaptation: maximizing yield stability over different
environment types.

\subsection{Acknowledgment}\label{acknowledgment}

This work benefited from the GENOPLANTE program \emph{HP1} (2001--2004),
the \emph{SUNYFUEL} project, financially supported by the French
National Research Agency (ANR-07-GPLA-0022, 2008--2011), the
\emph{OLEOSOL} project (2009--2012) with the financial support from the
Midi Pyrénées Region, the European Fund for Regional Development (EFRD),
and the French Fund for Competitiveness Clusters (FUI), and the
\emph{SUNRISE} project of the French National Research Agency
(ANR-11-BTBR-0005, 2012-2019).

\newpage

\textbf{Table 1 Details on location, treatment, year, testers and
observed hybrids for the 17 environments.} \textsuperscript{a} The
locations were designated as follows: AI: Aigrefeuille (Center West),
CA: Castelnaudary (South West), CO : Cornebarrieu (South West), GA:
Gaillac (South West), VE: Verdun (South West), LO: Loudun (Center West),
SE: Segoufielle (South West), CHA: Chateauroux (Center).
\textsuperscript{b} I for irrigation, NI for non irrigation

\begin{longtable}[]{@{}lllrllr@{}}
\toprule
\begin{minipage}[b]{0.10\columnwidth}\raggedright\strut
Environment\strut
\end{minipage} & \begin{minipage}[b]{0.10\columnwidth}\raggedright\strut
Location\textsuperscript{a}\strut
\end{minipage} & \begin{minipage}[b]{0.11\columnwidth}\raggedright\strut
Treatment\textsuperscript{b}\strut
\end{minipage} & \begin{minipage}[b]{0.05\columnwidth}\raggedleft\strut
Year\strut
\end{minipage} & \begin{minipage}[b]{0.15\columnwidth}\raggedright\strut
Tester for B-line\strut
\end{minipage} & \begin{minipage}[b]{0.15\columnwidth}\raggedright\strut
Tester for R-line\strut
\end{minipage} & \begin{minipage}[b]{0.14\columnwidth}\raggedleft\strut
Number of lines\strut
\end{minipage}\tabularnewline
\midrule
\endhead
\begin{minipage}[t]{0.10\columnwidth}\raggedright\strut
AI08\_I\strut
\end{minipage} & \begin{minipage}[t]{0.10\columnwidth}\raggedright\strut
AI\strut
\end{minipage} & \begin{minipage}[t]{0.11\columnwidth}\raggedright\strut
I\strut
\end{minipage} & \begin{minipage}[t]{0.05\columnwidth}\raggedleft\strut
2008\strut
\end{minipage} & \begin{minipage}[t]{0.15\columnwidth}\raggedright\strut
83HR4gms\strut
\end{minipage} & \begin{minipage}[t]{0.15\columnwidth}\raggedright\strut
T1\strut
\end{minipage} & \begin{minipage}[t]{0.14\columnwidth}\raggedleft\strut
192\strut
\end{minipage}\tabularnewline
\begin{minipage}[t]{0.10\columnwidth}\raggedright\strut
AI08\_NI\strut
\end{minipage} & \begin{minipage}[t]{0.10\columnwidth}\raggedright\strut
AI\strut
\end{minipage} & \begin{minipage}[t]{0.11\columnwidth}\raggedright\strut
NI\strut
\end{minipage} & \begin{minipage}[t]{0.05\columnwidth}\raggedleft\strut
2008\strut
\end{minipage} & \begin{minipage}[t]{0.15\columnwidth}\raggedright\strut
83HR4gms\strut
\end{minipage} & \begin{minipage}[t]{0.15\columnwidth}\raggedright\strut
T1\strut
\end{minipage} & \begin{minipage}[t]{0.14\columnwidth}\raggedleft\strut
193\strut
\end{minipage}\tabularnewline
\begin{minipage}[t]{0.10\columnwidth}\raggedright\strut
CO09\_I\strut
\end{minipage} & \begin{minipage}[t]{0.10\columnwidth}\raggedright\strut
CO\strut
\end{minipage} & \begin{minipage}[t]{0.11\columnwidth}\raggedright\strut
I\strut
\end{minipage} & \begin{minipage}[t]{0.05\columnwidth}\raggedleft\strut
2009\strut
\end{minipage} & \begin{minipage}[t]{0.15\columnwidth}\raggedright\strut
83HR4gms\strut
\end{minipage} & \begin{minipage}[t]{0.15\columnwidth}\raggedright\strut
T1\strut
\end{minipage} & \begin{minipage}[t]{0.14\columnwidth}\raggedleft\strut
278\strut
\end{minipage}\tabularnewline
\begin{minipage}[t]{0.10\columnwidth}\raggedright\strut
CO09\_NI\strut
\end{minipage} & \begin{minipage}[t]{0.10\columnwidth}\raggedright\strut
CO\strut
\end{minipage} & \begin{minipage}[t]{0.11\columnwidth}\raggedright\strut
NI\strut
\end{minipage} & \begin{minipage}[t]{0.05\columnwidth}\raggedleft\strut
2009\strut
\end{minipage} & \begin{minipage}[t]{0.15\columnwidth}\raggedright\strut
83HR4gms\strut
\end{minipage} & \begin{minipage}[t]{0.15\columnwidth}\raggedright\strut
T1\strut
\end{minipage} & \begin{minipage}[t]{0.14\columnwidth}\raggedleft\strut
278\strut
\end{minipage}\tabularnewline
\begin{minipage}[t]{0.10\columnwidth}\raggedright\strut
GA09\_I\strut
\end{minipage} & \begin{minipage}[t]{0.10\columnwidth}\raggedright\strut
GA\strut
\end{minipage} & \begin{minipage}[t]{0.11\columnwidth}\raggedright\strut
I\strut
\end{minipage} & \begin{minipage}[t]{0.05\columnwidth}\raggedleft\strut
2009\strut
\end{minipage} & \begin{minipage}[t]{0.15\columnwidth}\raggedright\strut
83HR4gms\strut
\end{minipage} & \begin{minipage}[t]{0.15\columnwidth}\raggedright\strut
T1\strut
\end{minipage} & \begin{minipage}[t]{0.14\columnwidth}\raggedleft\strut
275\strut
\end{minipage}\tabularnewline
\begin{minipage}[t]{0.10\columnwidth}\raggedright\strut
GA09\_NI\strut
\end{minipage} & \begin{minipage}[t]{0.10\columnwidth}\raggedright\strut
GA\strut
\end{minipage} & \begin{minipage}[t]{0.11\columnwidth}\raggedright\strut
NI\strut
\end{minipage} & \begin{minipage}[t]{0.05\columnwidth}\raggedleft\strut
2009\strut
\end{minipage} & \begin{minipage}[t]{0.15\columnwidth}\raggedright\strut
83HR4gms\strut
\end{minipage} & \begin{minipage}[t]{0.15\columnwidth}\raggedright\strut
T1\strut
\end{minipage} & \begin{minipage}[t]{0.14\columnwidth}\raggedleft\strut
274\strut
\end{minipage}\tabularnewline
\begin{minipage}[t]{0.10\columnwidth}\raggedright\strut
LO10\_NI\strut
\end{minipage} & \begin{minipage}[t]{0.10\columnwidth}\raggedright\strut
LO\strut
\end{minipage} & \begin{minipage}[t]{0.11\columnwidth}\raggedright\strut
NI\strut
\end{minipage} & \begin{minipage}[t]{0.05\columnwidth}\raggedleft\strut
2010\strut
\end{minipage} & \begin{minipage}[t]{0.15\columnwidth}\raggedright\strut
83HR4gms\strut
\end{minipage} & \begin{minipage}[t]{0.15\columnwidth}\raggedright\strut
T1\strut
\end{minipage} & \begin{minipage}[t]{0.14\columnwidth}\raggedleft\strut
284\strut
\end{minipage}\tabularnewline
\begin{minipage}[t]{0.10\columnwidth}\raggedright\strut
VE10\_I\strut
\end{minipage} & \begin{minipage}[t]{0.10\columnwidth}\raggedright\strut
VE\strut
\end{minipage} & \begin{minipage}[t]{0.11\columnwidth}\raggedright\strut
I\strut
\end{minipage} & \begin{minipage}[t]{0.05\columnwidth}\raggedleft\strut
2010\strut
\end{minipage} & \begin{minipage}[t]{0.15\columnwidth}\raggedright\strut
83HR4gms\strut
\end{minipage} & \begin{minipage}[t]{0.15\columnwidth}\raggedright\strut
T1\strut
\end{minipage} & \begin{minipage}[t]{0.14\columnwidth}\raggedleft\strut
289\strut
\end{minipage}\tabularnewline
\begin{minipage}[t]{0.10\columnwidth}\raggedright\strut
AI09\_I\strut
\end{minipage} & \begin{minipage}[t]{0.10\columnwidth}\raggedright\strut
AI\strut
\end{minipage} & \begin{minipage}[t]{0.11\columnwidth}\raggedright\strut
I\strut
\end{minipage} & \begin{minipage}[t]{0.05\columnwidth}\raggedleft\strut
2009\strut
\end{minipage} & \begin{minipage}[t]{0.15\columnwidth}\raggedright\strut
T2\strut
\end{minipage} & \begin{minipage}[t]{0.15\columnwidth}\raggedright\strut
T3\strut
\end{minipage} & \begin{minipage}[t]{0.14\columnwidth}\raggedleft\strut
280\strut
\end{minipage}\tabularnewline
\begin{minipage}[t]{0.10\columnwidth}\raggedright\strut
AI09\_NI\strut
\end{minipage} & \begin{minipage}[t]{0.10\columnwidth}\raggedright\strut
AI\strut
\end{minipage} & \begin{minipage}[t]{0.11\columnwidth}\raggedright\strut
NI\strut
\end{minipage} & \begin{minipage}[t]{0.05\columnwidth}\raggedleft\strut
2009\strut
\end{minipage} & \begin{minipage}[t]{0.15\columnwidth}\raggedright\strut
T2\strut
\end{minipage} & \begin{minipage}[t]{0.15\columnwidth}\raggedright\strut
T3\strut
\end{minipage} & \begin{minipage}[t]{0.14\columnwidth}\raggedleft\strut
280\strut
\end{minipage}\tabularnewline
\begin{minipage}[t]{0.10\columnwidth}\raggedright\strut
VE09\_I\strut
\end{minipage} & \begin{minipage}[t]{0.10\columnwidth}\raggedright\strut
VE\strut
\end{minipage} & \begin{minipage}[t]{0.11\columnwidth}\raggedright\strut
I\strut
\end{minipage} & \begin{minipage}[t]{0.05\columnwidth}\raggedleft\strut
2009\strut
\end{minipage} & \begin{minipage}[t]{0.15\columnwidth}\raggedright\strut
T2\strut
\end{minipage} & \begin{minipage}[t]{0.15\columnwidth}\raggedright\strut
T3\strut
\end{minipage} & \begin{minipage}[t]{0.14\columnwidth}\raggedleft\strut
273\strut
\end{minipage}\tabularnewline
\begin{minipage}[t]{0.10\columnwidth}\raggedright\strut
VE09\_NI\strut
\end{minipage} & \begin{minipage}[t]{0.10\columnwidth}\raggedright\strut
VE\strut
\end{minipage} & \begin{minipage}[t]{0.11\columnwidth}\raggedright\strut
NI\strut
\end{minipage} & \begin{minipage}[t]{0.05\columnwidth}\raggedleft\strut
2009\strut
\end{minipage} & \begin{minipage}[t]{0.15\columnwidth}\raggedright\strut
T2\strut
\end{minipage} & \begin{minipage}[t]{0.15\columnwidth}\raggedright\strut
T3\strut
\end{minipage} & \begin{minipage}[t]{0.14\columnwidth}\raggedleft\strut
273\strut
\end{minipage}\tabularnewline
\begin{minipage}[t]{0.10\columnwidth}\raggedright\strut
CA10\_NI\strut
\end{minipage} & \begin{minipage}[t]{0.10\columnwidth}\raggedright\strut
C1\strut
\end{minipage} & \begin{minipage}[t]{0.11\columnwidth}\raggedright\strut
NI\strut
\end{minipage} & \begin{minipage}[t]{0.05\columnwidth}\raggedleft\strut
2010\strut
\end{minipage} & \begin{minipage}[t]{0.15\columnwidth}\raggedright\strut
T2\strut
\end{minipage} & \begin{minipage}[t]{0.15\columnwidth}\raggedright\strut
T3\strut
\end{minipage} & \begin{minipage}[t]{0.14\columnwidth}\raggedleft\strut
306\strut
\end{minipage}\tabularnewline
\begin{minipage}[t]{0.10\columnwidth}\raggedright\strut
CO08\_I\strut
\end{minipage} & \begin{minipage}[t]{0.10\columnwidth}\raggedright\strut
CO\strut
\end{minipage} & \begin{minipage}[t]{0.11\columnwidth}\raggedright\strut
I\strut
\end{minipage} & \begin{minipage}[t]{0.05\columnwidth}\raggedleft\strut
2008\strut
\end{minipage} & \begin{minipage}[t]{0.15\columnwidth}\raggedright\strut
T2\strut
\end{minipage} & \begin{minipage}[t]{0.15\columnwidth}\raggedright\strut
T3\strut
\end{minipage} & \begin{minipage}[t]{0.14\columnwidth}\raggedleft\strut
249\strut
\end{minipage}\tabularnewline
\begin{minipage}[t]{0.10\columnwidth}\raggedright\strut
CO08\_NI\strut
\end{minipage} & \begin{minipage}[t]{0.10\columnwidth}\raggedright\strut
CO\strut
\end{minipage} & \begin{minipage}[t]{0.11\columnwidth}\raggedright\strut
NI\strut
\end{minipage} & \begin{minipage}[t]{0.05\columnwidth}\raggedleft\strut
2008\strut
\end{minipage} & \begin{minipage}[t]{0.15\columnwidth}\raggedright\strut
T2\strut
\end{minipage} & \begin{minipage}[t]{0.15\columnwidth}\raggedright\strut
T3\strut
\end{minipage} & \begin{minipage}[t]{0.14\columnwidth}\raggedleft\strut
249\strut
\end{minipage}\tabularnewline
\begin{minipage}[t]{0.10\columnwidth}\raggedright\strut
SE10\_NI\strut
\end{minipage} & \begin{minipage}[t]{0.10\columnwidth}\raggedright\strut
SE\strut
\end{minipage} & \begin{minipage}[t]{0.11\columnwidth}\raggedright\strut
NI\strut
\end{minipage} & \begin{minipage}[t]{0.05\columnwidth}\raggedleft\strut
2010\strut
\end{minipage} & \begin{minipage}[t]{0.15\columnwidth}\raggedright\strut
T2\strut
\end{minipage} & \begin{minipage}[t]{0.15\columnwidth}\raggedright\strut
T3\strut
\end{minipage} & \begin{minipage}[t]{0.14\columnwidth}\raggedleft\strut
285\strut
\end{minipage}\tabularnewline
\begin{minipage}[t]{0.10\columnwidth}\raggedright\strut
CHA10\_I\strut
\end{minipage} & \begin{minipage}[t]{0.10\columnwidth}\raggedright\strut
CHA\strut
\end{minipage} & \begin{minipage}[t]{0.11\columnwidth}\raggedright\strut
NI\strut
\end{minipage} & \begin{minipage}[t]{0.05\columnwidth}\raggedleft\strut
2010\strut
\end{minipage} & \begin{minipage}[t]{0.15\columnwidth}\raggedright\strut
T2\strut
\end{minipage} & \begin{minipage}[t]{0.15\columnwidth}\raggedright\strut
T3\strut
\end{minipage} & \begin{minipage}[t]{0.14\columnwidth}\raggedleft\strut
306\strut
\end{minipage}\tabularnewline
\bottomrule
\end{longtable}

\newpage

\textbf{Table 2: Description of abiotic stress indicators simulated by
the crop model.} The SUNFLO crop model was used to simulate the
interactions between plant growth and available environmental resources.
The evolution of resources level and abiotic constraints during the crop
cycle was summarized by computing 8 stress indicators during 7 cropping
periods: vegetative, flowering, grain filling phase and their
combination. \(1_{[x]}\) equals 1 if \(x\) is true and 0 else.

\begin{longtable}[]{@{}lllll@{}}
\toprule
\begin{minipage}[b]{0.10\columnwidth}\raggedright\strut
Stress\strut
\end{minipage} & \begin{minipage}[b]{0.06\columnwidth}\raggedright\strut
Symbol\strut
\end{minipage} & \begin{minipage}[b]{0.29\columnwidth}\raggedright\strut
Description\strut
\end{minipage} & \begin{minipage}[b]{0.09\columnwidth}\raggedright\strut
Unit\strut
\end{minipage} & \begin{minipage}[b]{0.32\columnwidth}\raggedright\strut
Formula\strut
\end{minipage}\tabularnewline
\midrule
\endhead
\begin{minipage}[t]{0.10\columnwidth}\raggedright\strut
temperature\strut
\end{minipage} & \begin{minipage}[t]{0.06\columnwidth}\raggedright\strut
HTi\strut
\end{minipage} & \begin{minipage}[t]{0.29\columnwidth}\raggedright\strut
high temperature (continuous)\strut
\end{minipage} & \begin{minipage}[t]{0.09\columnwidth}\raggedright\strut
-\strut
\end{minipage} & \begin{minipage}[t]{0.32\columnwidth}\raggedright\strut
\(\int 1-\textrm{HTRUE} \, \mathrm{d}t\)\strut
\end{minipage}\tabularnewline
\begin{minipage}[t]{0.10\columnwidth}\raggedright\strut
temperature\strut
\end{minipage} & \begin{minipage}[t]{0.06\columnwidth}\raggedright\strut
HTs\strut
\end{minipage} & \begin{minipage}[t]{0.29\columnwidth}\raggedright\strut
high temperature (discrete)\strut
\end{minipage} & \begin{minipage}[t]{0.09\columnwidth}\raggedright\strut
d\strut
\end{minipage} & \begin{minipage}[t]{0.32\columnwidth}\raggedright\strut
\(\sum 1_{[T_m > 28]}\)\strut
\end{minipage}\tabularnewline
\begin{minipage}[t]{0.10\columnwidth}\raggedright\strut
temperature\strut
\end{minipage} & \begin{minipage}[t]{0.06\columnwidth}\raggedright\strut
LTi\strut
\end{minipage} & \begin{minipage}[t]{0.29\columnwidth}\raggedright\strut
low temperature (continuous)\strut
\end{minipage} & \begin{minipage}[t]{0.09\columnwidth}\raggedright\strut
-\strut
\end{minipage} & \begin{minipage}[t]{0.32\columnwidth}\raggedright\strut
\(\int 1-\textrm{LTRUE} \, \mathrm{d}t\)\strut
\end{minipage}\tabularnewline
\begin{minipage}[t]{0.10\columnwidth}\raggedright\strut
temperature\strut
\end{minipage} & \begin{minipage}[t]{0.06\columnwidth}\raggedright\strut
LTs\strut
\end{minipage} & \begin{minipage}[t]{0.29\columnwidth}\raggedright\strut
low temperature (discrete)\strut
\end{minipage} & \begin{minipage}[t]{0.09\columnwidth}\raggedright\strut
d\strut
\end{minipage} & \begin{minipage}[t]{0.32\columnwidth}\raggedright\strut
\(\sum 1_{[T_m < 20]}\)\strut
\end{minipage}\tabularnewline
\begin{minipage}[t]{0.10\columnwidth}\raggedright\strut
water\strut
\end{minipage} & \begin{minipage}[t]{0.06\columnwidth}\raggedright\strut
FTSW\strut
\end{minipage} & \begin{minipage}[t]{0.29\columnwidth}\raggedright\strut
Edaphic water deficit (continuous)\strut
\end{minipage} & \begin{minipage}[t]{0.09\columnwidth}\raggedright\strut
-\strut
\end{minipage} & \begin{minipage}[t]{0.32\columnwidth}\raggedright\strut
\(\int 1-\textrm{FTSW} \, \mathrm{d}t\)\strut
\end{minipage}\tabularnewline
\begin{minipage}[t]{0.10\columnwidth}\raggedright\strut
water\strut
\end{minipage} & \begin{minipage}[t]{0.06\columnwidth}\raggedright\strut
ETR\strut
\end{minipage} & \begin{minipage}[t]{0.29\columnwidth}\raggedright\strut
Edaphic water deficit (discrete)\strut
\end{minipage} & \begin{minipage}[t]{0.09\columnwidth}\raggedright\strut
d\strut
\end{minipage} & \begin{minipage}[t]{0.32\columnwidth}\raggedright\strut
\(\sum 1_{[\textrm{ET/PET} < 0.6]}\)\strut
\end{minipage}\tabularnewline
\begin{minipage}[t]{0.10\columnwidth}\raggedright\strut
nitrogen\strut
\end{minipage} & \begin{minipage}[t]{0.06\columnwidth}\raggedright\strut
NAB\strut
\end{minipage} & \begin{minipage}[t]{0.29\columnwidth}\raggedright\strut
Absorbed nitrogen\strut
\end{minipage} & \begin{minipage}[t]{0.09\columnwidth}\raggedright\strut
kg ha\textsuperscript{-1}\strut
\end{minipage} & \begin{minipage}[t]{0.32\columnwidth}\raggedright\strut
\(\int \textrm{NAB} \, \mathrm{d}t\)\strut
\end{minipage}\tabularnewline
\begin{minipage}[t]{0.10\columnwidth}\raggedright\strut
nitrogen\strut
\end{minipage} & \begin{minipage}[t]{0.06\columnwidth}\raggedright\strut
NNI\strut
\end{minipage} & \begin{minipage}[t]{0.29\columnwidth}\raggedright\strut
Nitrogen deficit (continuous)\strut
\end{minipage} & \begin{minipage}[t]{0.09\columnwidth}\raggedright\strut
-\strut
\end{minipage} & \begin{minipage}[t]{0.32\columnwidth}\raggedright\strut
\(\int 1-\textrm{NNI} \, \mathrm{d}t\)\strut
\end{minipage}\tabularnewline
\bottomrule
\end{longtable}

\newpage

\textbf{Table 3. Results of the regression between grain yield and
abiotic environmental indicators: cold, nitrogen and water stresses.}
Plasticities of the three control genotypes (i.e.~commercial varieties)
are presented as their slopes and p-values of the type III Fisher test
in the best model with the proportion of explained variance for each
control genotype. \textsuperscript{a} LTi\_veg: integration of low
temperatures during the vegetative stage, \textsuperscript{b} NAB:
absorbed nitrogen during the whole growth period \textsuperscript{c}
FTSW\_veg+flo: fraction of the transpirable soil water during the
vegetative and the flowering periods. The best explanatory linear model
was found by AIC criterion among all the regression models having one to
four linear regressors and at most one indicator per stress.

\begin{tabular}{l|l|l|rrr}
\hline
  &   & Unit & Melody & Pacific & Pegasol\\
\hline
Yield &  & q ha$^{-1}$ & 32.41 & 29.18 & 31.30\\
Cold stress$^a$ & slope & q ha$^{-1}$ d$^{-1}$ & -0.36 & -0.36 & -0.32\\
Cold stress & p-value &  & 2.39 10$^{-4}$ & 1.97 10$^{-3}$ & 1.95 10$^{-3}$\\
Nitrogen stress$^b$ & slope & q ha$^{-1}$ kg$^{-1}$ & 0.27 & 0.22 & 0.25\\
Nitrogen stress & p-value &  & 1.16 10$^{-3}$ & 7.66 10$^{-3}$ & 6.6 10$^{-4}$\\
Water stress$^c$ & slope & q ha$^{-1}$ d$^{-1}$ & -0.16 & -0.12 & -0.13\\
Water stress & p-value &  & 3.27 10$^{-3}$ & 2.91 10$^{-2}$& 8.21 10$^{-3}$\\
Explained variance  &  &  & 0.93 & 0.94 & 0.90\\
\hline
\end{tabular}

\newpage

\textbf{Table 4. Variation and correlation of the plasticities of oil
yield for different abiotic stresses in the diversity panel.} Minimum,
mean, maximum, and variance of the plasticity phenotypes and their
correlations. The plasticity phenotypes are calculated as the slopes in
a linear model including the three stress indicators that best
characterized the environments. \textsuperscript{a} cold stress
plasticity (q h\textsuperscript{-1} d\textsuperscript{-1}),
\textsuperscript{b} nitrogen stress plasticity (q h\textsuperscript{-1}
kg\textsuperscript{-1}), \textsuperscript{c} water stress plasticity (q
h\textsuperscript{-1} d\textsuperscript{-1})

\begin{tabular}{l|rrr|r|r|r}
\hline
  & Min.  & Mean & Max. & Var. & \multicolumn{2}{c}{Cor.}\\
 & & & & & Nitrogen & Water \\
\hline
Cold stress$^a$ & -0.37 & -0.23 & -0.09 & 2.5 10$^{-3}$ & 0.12 & 0.03 \\
Nitrogen stress$^b$ & 0.01 & 0.11 & 0.21 & 8.1 10$^{-4}$ & & 0.62\\
Water stress$^c$ & -0.25 & -0.08 & 0.16 & 3.7 10$^{-3}$ \\
\hline
\end{tabular}

\newpage

\landscapebegin

\textbf{Table 5: Positions and estimated effects of SNP and genes
associated with oil yield plasticity to cold stress.} Genetic distance
determined on a consensus map built from three RIL populations (see
details in Materials and Methods). Physical position was determined via
BLASTing against the genome of line XRQ (early access to version
HanXRQv1.1). Association tests were performed using the MLMM procedure
unless otherwise stated and either with or without the
maintainer/restorer status structure (noted as B/R status). Effects of
SNP were estimated using a linear model including all associated SNP.
Their relative effect compared with the average plasticity of the
diversity panel is indicated. Sunflower gene names correspond to the
genome HannXRQ v1.1 (early access). The closest \emph{Arabidopsis
thaliana} homologue is indicated with its TAIR code, name, and a brief
functional description.

\tiny

\begin{tabular}{c|c|c|c|c|c|c|c|c}
\hline
SNP name & QTL & LG & Genetic & Physical & Association & Association model & Effect & Effect \\
    &       & & distance (cM) & position (bp) & test p-value &  & on plasticity & percentage \\
\hline
AX-84511295 & LG05.17 & 5 & 16.6 & 14,284,297 & 1.54 10$^{-8}$ & Without B/R status & 2.80 10${-2}$ & 12\% \\
AX-84248033 & LG05.64 & 5 & 64.4 & 206,893,338 & 5.70 10$^{-6}$ & EMMA with B/R status & 3.15 10$^{-2}$ & 13\% \\
 &  &  &  &  & 3.64 10$^{-6}$ & MLMM with B/R status & 3.34 10$^{-2}$ & 14\%  \\
AX-84488969 & LG09.27 & 9 & 27.4 & 203,648,151 & 6.76 10$^{-7}$ & With B/R status & -2.35 10$^{-2}$ & -10\% \\
AX-84358846 & LG10.34 & 10 & 34.1 & 220,548,129 & 3.87 10$^{-6}$ & Without B/R status & -5.06 10$^{-2}$ & -21\% \\
AX-84316040 & LG10.45 & 10 & 44.8 & 93,526,595 & 5.11 10$^{-7}$ & With B/R status & -3.14 10$^{-2}$ & -13\% \\
 &  &  &  & or 217,685,071 &  &  &  &  \\
AX-84586324 & LG13.72 & 13 & 72.4 & 190,712,390 & 1.17 10$^{-7}$ & Without B/R status & -2.94 10$^{-2}$ & -12\% \\
AX-84436021 & LG14.27 & 14 & 27.0 & 168,814,054 & 2.70 10$^{-7}$ & With B/R status & 3.61 10$^{-2}$ & 15\% \\
AX-84337313 & LG16.48 & 16 & 47.7 & 129,668,002 & 6.65 10$^{-8}$ & Without B/R status & -2.72 10$^{-2}$ & -12\% \\
AX-84507515 & LG17.49 & 17 & 49.1 & 153,030,442 & 6.81 10$^{-6}$ & EMMA without B/R status & 3.55 10$^{-2}$ & 15\% \\
\hline
\end{tabular}

(Table continued\ldots{})

\begin{tabular}{c|c|c|c|c}
\hline
SNP name & Sunflower gene & TAIR code & Description & Arabidopsis name\\
\hline
AX-84511295 & HannXRQ\_Chr05g0131961 & AFS71501G68570 & Oligopeptide transporter & NPF3.1\\
AX-84248033 & HannXRQ\_Chr05g0159711 & AT4G26720 & Protein phosphatase & PPX1\\
 & & & & \\
AX-84488969 & HannXRQ\_Chr09g0274811 & AT4G33495 & Root development, Temperature sensitive & RPD1\\
AX-84358846 & HannXRQ\_Chr10g0311881 & AT3G14150 & ROS protection & HAOX2\\
AX-84316040 & HannXRQ\_Chr10g0292461 & AT5G14950 & N-glycan mannose hyper-osmotic salinity stress & GMII\\
 & or HannXRQ\_Chr10g0311021 & & & \\
AX-84586324  & HannXRQ\_Chr13g0424331 & AT3G12490 & Cystatin, protease inhibitor, Tolerance to abiotic stresses incl. cold & CYS6\\
AX-84436021 & HannXRQ\_Chr14g0460351 & AT5G46040 & Oligopeptide transporter & NPF5.3\\
AX-84337313 & HannXRQ\_Chr16g0519611 & AT3G07450 & Lipid Transfer Protein & \\
AX-84507515 & HannXRQ\_Chr04g0119921 & AG24180 & Root development, Auxin & IAR4\\
\hline
\end{tabular}

\normalsize

\landscapeend

\newpage

\includegraphics{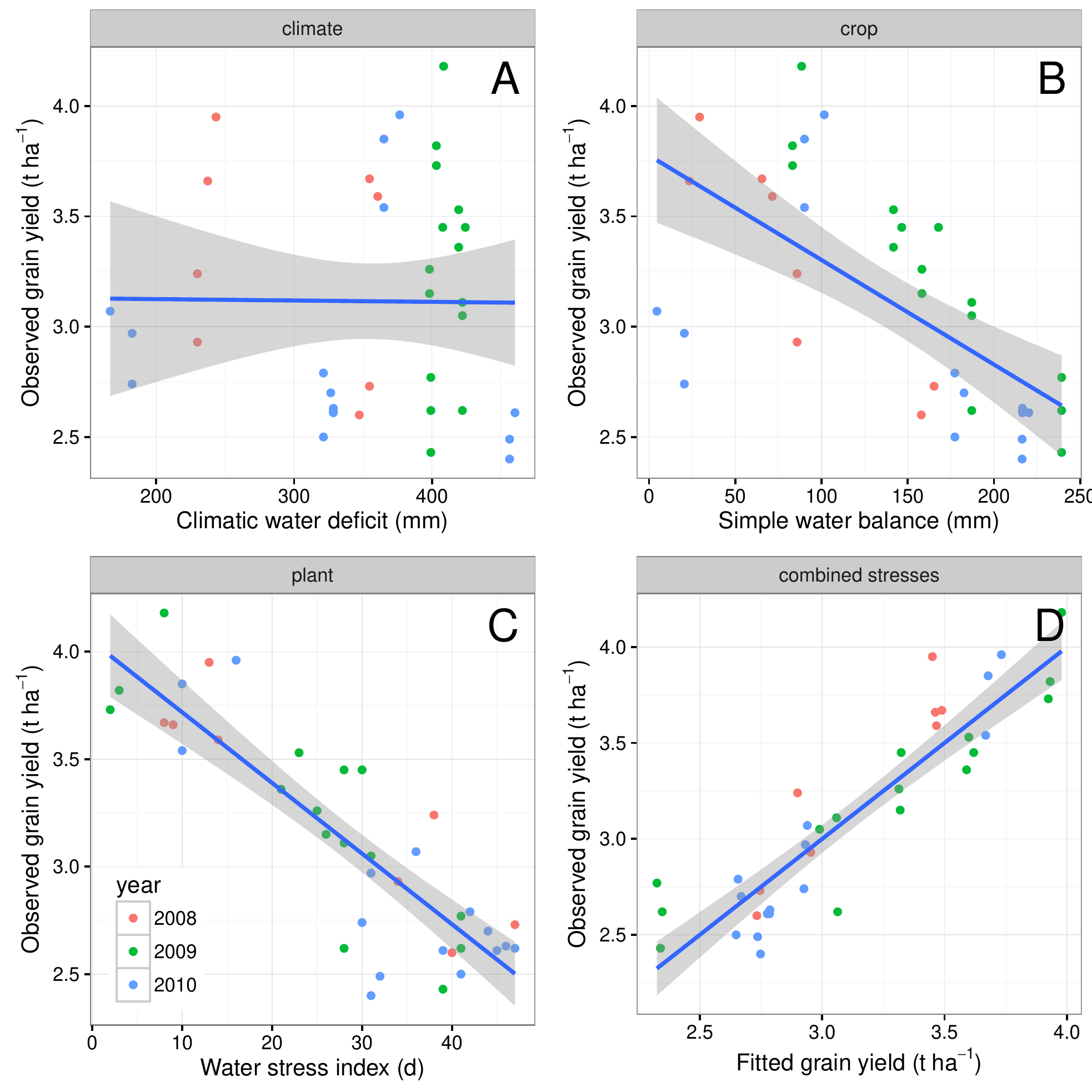}

\begin{quote}
\textbf{Figure 1. Relation between observed grain yield and several
abiotic stress indicators.} Panels A, B, and C display the regression
line between grain yield and water stress indicators, computed at
different levels: using climatic data only (panel A, precipitation -
potential evapotranspiration, Pearson correlation (\(r\)) of -0.01),
using both climatic and crop data (panel B, precipitation - potential
evapotranspiration + irrigation + soil water capacity, \(r\)) = -0.65,
p-value = 9.6 10\textsuperscript{-6}) and using simulated plant data
(panel C, evapotranspiration ratio, \(r\) = -0.86, p-value = 3.3
10\textsuperscript{-12}). Panel D displays grain yield predicted as a
function of combined abiotic stresses (linear model of water, nitrogen,
and cold stress indicators, \(r\) = 0.91, root mean square error of 0.21
t ha\textsuperscript{-1}).
\end{quote}

\newpage

\includegraphics{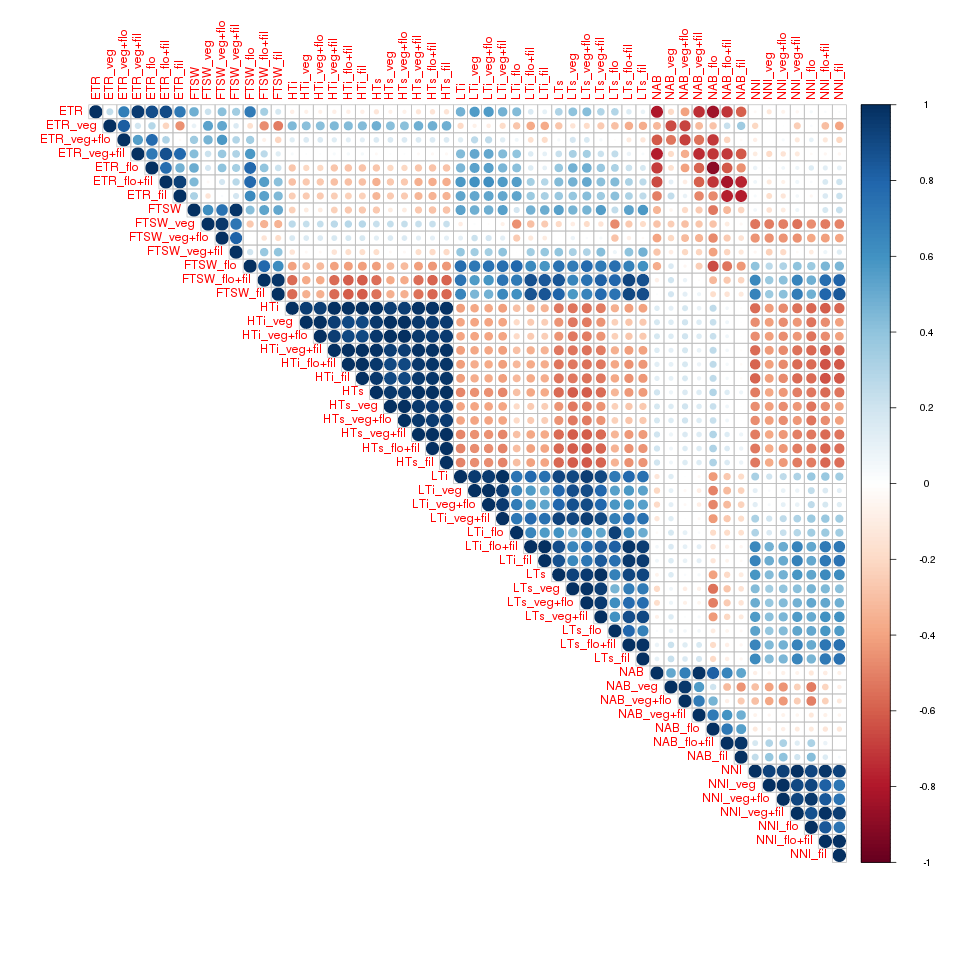}

\begin{quote}
\textbf{Figure 2. Heat map of the correlations between stress
indicators.} The three selected stress indicators are indicated in blue,
yellow and green for water, cold and nitrogen respectively. For water
stress, the fraction of transpirable soil water (FTSW) represent yield
limitation through water deficit (integration of 1 minus FTSW); ETR is
the conditional sum of days, if the ratio of the real evapotranspiration
(ET) to potential evapotranspiration (PET) was less than 0.6 (threshold
for photosynthesis limitation). For cold stress, LTs is the conditional
sum of days if mean air temperature was below 20°C and LTi represent low
temperatures impact on photosynthesis (integration of 1 minus equation
2). Heat stress indicators were computed following the same logic,
albeit representing high temperatures impact on photosynthesis.
Equations (2) and (3) are used in the crop model to define the radiation
use efficiency (RUE) response to temperature (Villalobos et al., 1996).
For nitrogen deficit, NAB is the amount of absorbed nitrogen in the
considered cropping period and NNI is the sum of 1 minus nitrogen
nutrition index, which indicates crop nitrogen deficit (Debaeke et al.,
2012; Lemaire and Meynard, 1997).
\end{quote}

\newpage

\includegraphics{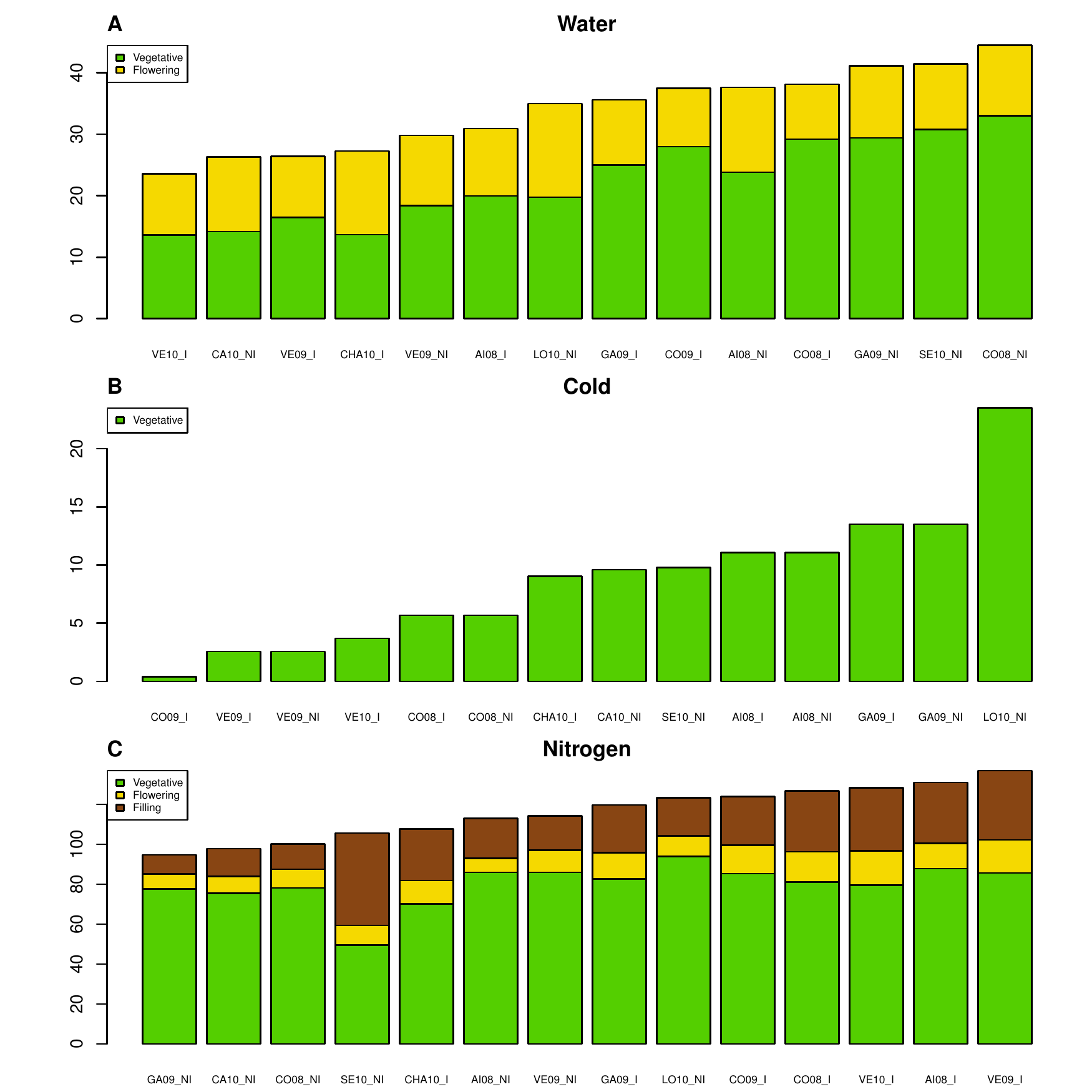}

\begin{quote}
\textbf{Figure 3. Range of variation in abiotic stresses in the
multi-environment trial.} Water (A), cold (B), and nitrogen stress (C)
indicators, computed by the SUNFLO crop model for each environment, and
averaged over the three control genotypes. The SUNFLO crop model was
used to compute stress indicators for three control commercial hybrids
(Melody, Pacific, Pegasol) according to the developmental stage of the
crop: vegetative growth (green), flowering (yellow), and seed filling
(brown).
\end{quote}

\newpage

\includegraphics{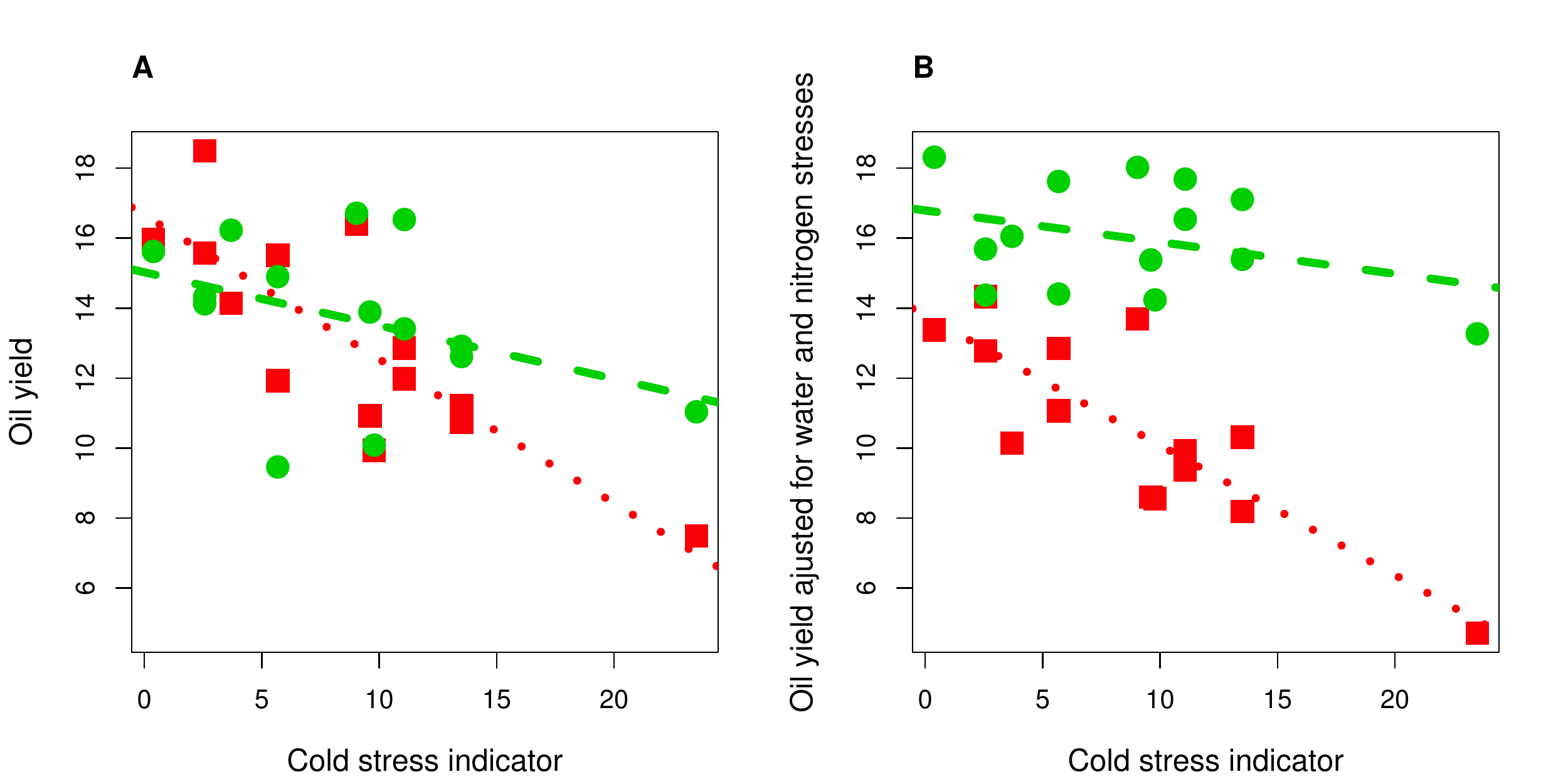}

\begin{quote}
\textbf{Figure 4. Regression of oil yield against the cold stress
indicator for the most tolerant (green triangles) and sensitive (red
squares) panel lines showing the ability of multi-stress modelling to
better characterize the environment.} A) Regression in a single (cold)
stress indicator model. B) Regression in a multi-stress (cold, nitrogen,
water) indicator model.
\end{quote}

\newpage

\includegraphics{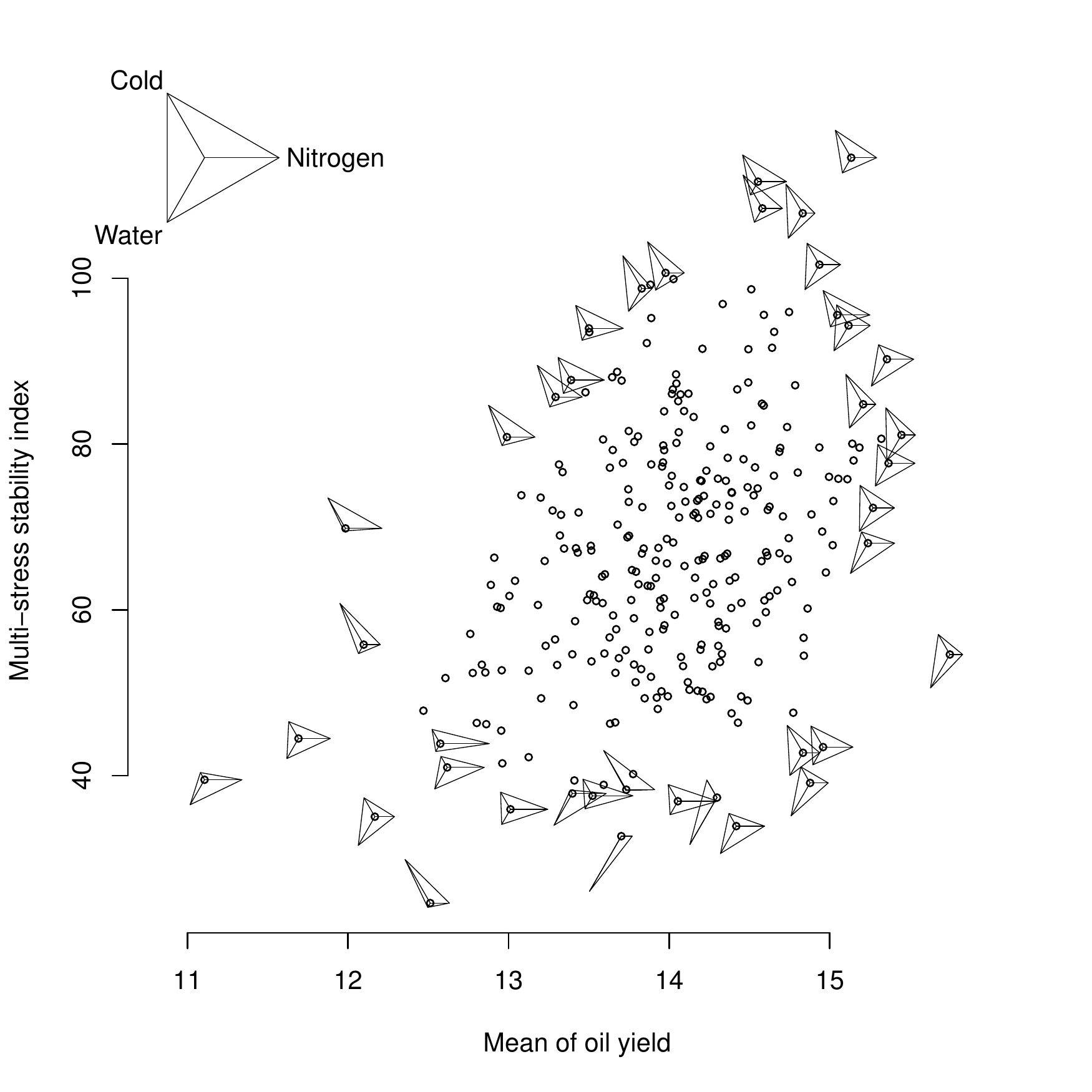}

\begin{quote}
\textbf{Figure 5. Relation between the multi-stress plasticity index and
the average oil yield showing differences in stress tolerance strategies
in the diversity panel of sunflower lines.} The three-branch star
represents the strategy of the most extreme panel lines in response to
combined abiotic stresses. The length of star branches represents the
relative plasticity against the corresponding stress: longer equals more
sensitive.
\end{quote}

\newpage

\subsection{Supplementary data}\label{supplementary-data}

\begin{longtable}[]{@{}lrrrrllr@{}}
\toprule
location & year & CWD & AWC & density & sowing & harvest &
irrigation\tabularnewline
\midrule
\endhead
AI08\_I & 2008 & -239.3 & 144.0 & 6.9 & May-06 & Sep-23 &
70.0\tabularnewline
AI08\_NI & 2008 & -229.8 & 144.0 & 6.9 & May-06 & Sep-16 &
0.0\tabularnewline
AI09\_I & 2009 & -408.5 & 240.0 & 6.9 & Apr-24 & Sep-07 &
80.0\tabularnewline
AI09\_NI & 2009 & -408.5 & 240.0 & 6.9 & Apr-24 & Sep-07 &
0.0\tabularnewline
CA10\_NI & 2010 & -323.1 & 144.0 & 6.6 & Apr-24 & Sep-11 &
0.0\tabularnewline
CHA10\_I & 2010 & -177.2 & 112.0 & 6.6 & Apr-30 & Oct-02 &
50.0\tabularnewline
CO08\_I & 2008 & -356.7 & 190.7 & 6.5 & May-22 & Oct-08 &
98.5\tabularnewline
CO08\_NI & 2008 & -349.8 & 189.4 & 6.5 & May-22 & Sep-19 &
0.0\tabularnewline
CO09\_I & 2009 & -421.0 & 174.7 & 6.5 & May-19 & Sep-29 &
103.0\tabularnewline
CO09\_NI & 2009 & -416.4 & 159.7 & 6.5 & May-19 & Sep-29 &
0.0\tabularnewline
GA09\_I & 2009 & -422.1 & 160.0 & 6.6 & May-06 & Sep-09 &
75.0\tabularnewline
GA09\_NI & 2009 & -399.2 & 160.0 & 6.6 & May-06 & Sep-02 &
0.0\tabularnewline
LO10\_NI & 2010 & -328.5 & 112.0 & 6.9 & Apr-15 & Sep-11 &
0.0\tabularnewline
SE10\_NI & 2010 & -457.6 & 240.0 & 6.5 & Apr-29 & Sep-30 &
0.0\tabularnewline
VE09\_I & 2009 & -405.0 & 240.0 & 6.9 & May-07 & Sep-10 &
80.0\tabularnewline
VE09\_NI & 2009 & -401.4 & 240.0 & 6.9 & May-07 & Sep-10 &
0.0\tabularnewline
VE10\_I & 2010 & -369.0 & 240.0 & 6.9 & May-07 & Sep-10 &
35.0\tabularnewline
\bottomrule
\end{longtable}

\begin{quote}
\textbf{Table S1. Description of locations and management practices on
the multi-environment trial.} Headers indicates the locations and years
of trials, the climatic water deficit (CWD, mm) i.e.\textasciitilde{}the
sum of precipitation minus sum of potential evapotranspiration, the
plant available water capacity (AWC, mm) i.e., the amount of soil water
reserves, plant density at sowing (plants m\textsuperscript{-2}), sowing
and harvest dates, and the amount of irrigation (mm).
\end{quote}

\newpage

\begin{longtable}[]{@{}llrrr@{}}
\toprule
\begin{minipage}[b]{0.55\columnwidth}\raggedright\strut
Parameter\strut
\end{minipage} & \begin{minipage}[b]{0.06\columnwidth}\raggedright\strut
Unit\strut
\end{minipage} & \begin{minipage}[b]{0.08\columnwidth}\raggedleft\strut
Melody\strut
\end{minipage} & \begin{minipage}[b]{0.09\columnwidth}\raggedleft\strut
Pacific\strut
\end{minipage} & \begin{minipage}[b]{0.09\columnwidth}\raggedleft\strut
Pegasol\strut
\end{minipage}\tabularnewline
\midrule
\endhead
\begin{minipage}[t]{0.55\columnwidth}\raggedright\strut
Temperature sum to floral initiation\strut
\end{minipage} & \begin{minipage}[t]{0.06\columnwidth}\raggedright\strut
°C d\strut
\end{minipage} & \begin{minipage}[t]{0.08\columnwidth}\raggedleft\strut
542\strut
\end{minipage} & \begin{minipage}[t]{0.09\columnwidth}\raggedleft\strut
531\strut
\end{minipage} & \begin{minipage}[t]{0.09\columnwidth}\raggedleft\strut
522\strut
\end{minipage}\tabularnewline
\begin{minipage}[t]{0.55\columnwidth}\raggedright\strut
Temperature sum from emergence to the beginning of flowering\strut
\end{minipage} & \begin{minipage}[t]{0.06\columnwidth}\raggedright\strut
°C d\strut
\end{minipage} & \begin{minipage}[t]{0.08\columnwidth}\raggedleft\strut
941\strut
\end{minipage} & \begin{minipage}[t]{0.09\columnwidth}\raggedleft\strut
922\strut
\end{minipage} & \begin{minipage}[t]{0.09\columnwidth}\raggedleft\strut
906\strut
\end{minipage}\tabularnewline
\begin{minipage}[t]{0.55\columnwidth}\raggedright\strut
Temperature sum from emergence to the beginning of grain filling\strut
\end{minipage} & \begin{minipage}[t]{0.06\columnwidth}\raggedright\strut
°C d\strut
\end{minipage} & \begin{minipage}[t]{0.08\columnwidth}\raggedleft\strut
1,188\strut
\end{minipage} & \begin{minipage}[t]{0.09\columnwidth}\raggedleft\strut
1,169\strut
\end{minipage} & \begin{minipage}[t]{0.09\columnwidth}\raggedleft\strut
1,153\strut
\end{minipage}\tabularnewline
\begin{minipage}[t]{0.55\columnwidth}\raggedright\strut
Temperature sum from emergence to seed physiological maturity\strut
\end{minipage} & \begin{minipage}[t]{0.06\columnwidth}\raggedright\strut
°C d\strut
\end{minipage} & \begin{minipage}[t]{0.08\columnwidth}\raggedleft\strut
1,751\strut
\end{minipage} & \begin{minipage}[t]{0.09\columnwidth}\raggedleft\strut
1,722\strut
\end{minipage} & \begin{minipage}[t]{0.09\columnwidth}\raggedleft\strut
1,721\strut
\end{minipage}\tabularnewline
\begin{minipage}[t]{0.55\columnwidth}\raggedright\strut
Potential number of leaves at flowering\strut
\end{minipage} & \begin{minipage}[t]{0.06\columnwidth}\raggedright\strut
leaf\strut
\end{minipage} & \begin{minipage}[t]{0.08\columnwidth}\raggedleft\strut
28.7\strut
\end{minipage} & \begin{minipage}[t]{0.09\columnwidth}\raggedleft\strut
23.5\strut
\end{minipage} & \begin{minipage}[t]{0.09\columnwidth}\raggedleft\strut
25.3\strut
\end{minipage}\tabularnewline
\begin{minipage}[t]{0.55\columnwidth}\raggedright\strut
Potential rank of the plant largest leaf at flowering\strut
\end{minipage} & \begin{minipage}[t]{0.06\columnwidth}\raggedright\strut
leaf\strut
\end{minipage} & \begin{minipage}[t]{0.08\columnwidth}\raggedleft\strut
17.4\strut
\end{minipage} & \begin{minipage}[t]{0.09\columnwidth}\raggedleft\strut
17.5\strut
\end{minipage} & \begin{minipage}[t]{0.09\columnwidth}\raggedleft\strut
25.3\strut
\end{minipage}\tabularnewline
\begin{minipage}[t]{0.55\columnwidth}\raggedright\strut
Potential area of the plant largest leaf at flowering\strut
\end{minipage} & \begin{minipage}[t]{0.06\columnwidth}\raggedright\strut
cm\textsuperscript{2}\strut
\end{minipage} & \begin{minipage}[t]{0.08\columnwidth}\raggedleft\strut
537\strut
\end{minipage} & \begin{minipage}[t]{0.09\columnwidth}\raggedleft\strut
420\strut
\end{minipage} & \begin{minipage}[t]{0.09\columnwidth}\raggedleft\strut
17.4\strut
\end{minipage}\tabularnewline
\begin{minipage}[t]{0.55\columnwidth}\raggedright\strut
Light extinction coefficient during vegetative growth\strut
\end{minipage} & \begin{minipage}[t]{0.06\columnwidth}\raggedright\strut
-\strut
\end{minipage} & \begin{minipage}[t]{0.08\columnwidth}\raggedleft\strut
0.838\strut
\end{minipage} & \begin{minipage}[t]{0.09\columnwidth}\raggedleft\strut
0.847\strut
\end{minipage} & \begin{minipage}[t]{0.09\columnwidth}\raggedleft\strut
0.856\strut
\end{minipage}\tabularnewline
\begin{minipage}[t]{0.55\columnwidth}\raggedright\strut
Threshold for leaf expansion response to water stress\strut
\end{minipage} & \begin{minipage}[t]{0.06\columnwidth}\raggedright\strut
-\strut
\end{minipage} & \begin{minipage}[t]{0.08\columnwidth}\raggedleft\strut
-3.896\strut
\end{minipage} & \begin{minipage}[t]{0.09\columnwidth}\raggedleft\strut
-3.359\strut
\end{minipage} & \begin{minipage}[t]{0.09\columnwidth}\raggedleft\strut
-3.687\strut
\end{minipage}\tabularnewline
\begin{minipage}[t]{0.55\columnwidth}\raggedright\strut
Threshold for stomatal conductance response to water stress\strut
\end{minipage} & \begin{minipage}[t]{0.06\columnwidth}\raggedright\strut
-\strut
\end{minipage} & \begin{minipage}[t]{0.08\columnwidth}\raggedleft\strut
-10.7\strut
\end{minipage} & \begin{minipage}[t]{0.09\columnwidth}\raggedleft\strut
-10.12\strut
\end{minipage} & \begin{minipage}[t]{0.09\columnwidth}\raggedleft\strut
-9.998\strut
\end{minipage}\tabularnewline
\begin{minipage}[t]{0.55\columnwidth}\raggedright\strut
Potential harvest index\strut
\end{minipage} & \begin{minipage}[t]{0.06\columnwidth}\raggedright\strut
-\strut
\end{minipage} & \begin{minipage}[t]{0.08\columnwidth}\raggedleft\strut
0.42\strut
\end{minipage} & \begin{minipage}[t]{0.09\columnwidth}\raggedleft\strut
0.39\strut
\end{minipage} & \begin{minipage}[t]{0.09\columnwidth}\raggedleft\strut
0.44\strut
\end{minipage}\tabularnewline
\begin{minipage}[t]{0.55\columnwidth}\raggedright\strut
Potential seed oil content\strut
\end{minipage} & \begin{minipage}[t]{0.06\columnwidth}\raggedright\strut
\% dry mass\strut
\end{minipage} & \begin{minipage}[t]{0.08\columnwidth}\raggedleft\strut
45.6\strut
\end{minipage} & \begin{minipage}[t]{0.09\columnwidth}\raggedleft\strut
46.5\strut
\end{minipage} & \begin{minipage}[t]{0.09\columnwidth}\raggedleft\strut
47.3\strut
\end{minipage}\tabularnewline
\bottomrule
\end{longtable}

\begin{quote}
\textbf{Table S2. SUNFLO genotype-dependent parameters for the three
controls.}
\end{quote}

\newpage

\footnotesize

\begin{tabular}{l|lll|lll|lll|l}
\hline
& \multicolumn{3}{c}{INEDI XRQxPSC8} & \multicolumn{3}{c}{FUxPAZ2} & \multicolumn{3}{c}{RHA801xRHA280 on consensus} & LD mapping \\
 & \# markers & \# pos. & cM &  \# markers & \# pos. &  cM & \# markers &  \# pos. &  cM & \# markers\\
\hline
LG1 & 2,585 & 118 & 72.2 & 949 & 45 & 61.0 & 4,474 & 297 & 60.5 & 5,350\\
LG2 & 1,356 & 98 & 78.8 & 726 & 35 & 68.4 & 2,727 & 229 & 73.2 & 3,347\\
LG3 & 1,907 & 105 & 90.3 & 1,217 & 62 & 91.0 & 4,904 & 307 & 84.1 & 5,654\\
LG4 & 1,781 & 118 & 108.7 & 1,428 & 70 & 115.9 & 4,644 & 329 & 106.8 & 5,303\\
LG5 & 2,100 & 112 & 87.6 & 1,133 & 46 & 111.5 & 4,982 & 259 & 99.0 & 5,933\\
LG6 & 2,583 & 113 & 56.3 & 853 & 30 & 47.5 & 4,152 & 255 & 55.3 & 4,950\\
LG7 & 1,280 & 80 & 71.0 & 682 & 36 & 79.5 & 2,432 & 221 & 68.4 & 3,145\\
LG8 & 2,069 & 122 & 67.7 & 959 & 47 & 64.4 & 4,550 & 298 & 65.2 & 5,417\\
LG9 & 2,508 & 126 & 106.4 & 1,219 & 57 & 99.6 & 5,980 & 346 & 94.4 & 6,893\\
LG10 & 2,880 & 126 & 102.3 & 968 & 50 & 89.8 & 6,314 & 335 & 90.5 & 7,004\\
LG11 & 1,922 & 129 & 94.3 & 643 & 37 & 100.2 & 3,708 & 268 & 87.6 & 4,706\\
LG12 & 654 & 60 & 79.7 & 673 & 35 & 58.7 & 3,321 & 219 & 65.7 & 4,398\\
LG13 & 759 & 88 & 75.1 & 1,096 & 44 & 83.9 & 3,578 & 237 & 77.5 & 4,253\\
LG14 & 2,585 & 91 & 77.7 & 451 & 23 & 50.2 & 5,235 & 212 & 74.3 & 6,432\\
LG15 & 1,486 & 97 & 99.7 & 1,097 & 46 & 112.4 & 4,004 & 259 & 92.8 & 4,940\\
LG16 & 2,672 & 169 & 109.7 & 1,367 & 71 & 89.8 & 5,100 & 405 & 92.4 & 6,061\\
LG17 & 630 & 109 & 110.2 & 2,440 & 73 & 101.5 & 5,045 & 306 & 110.9 & 6,193\\
\hline
map & 31,757 & 1,861 & 1,487.7 & 17,901 & 807 & 1,425.3 & 75,150 & 4,782 & 1,398.5 & 89,979\\
\hline
\end{tabular}

\normalsize

\begin{quote}
\textbf{Table S3. Marker number, genetic positions (pos.) and genetic
distance (cM) of the genetic maps and number of markers positionned by
linkage disequilibrium (LD).}
\end{quote}

\newpage

\includegraphics{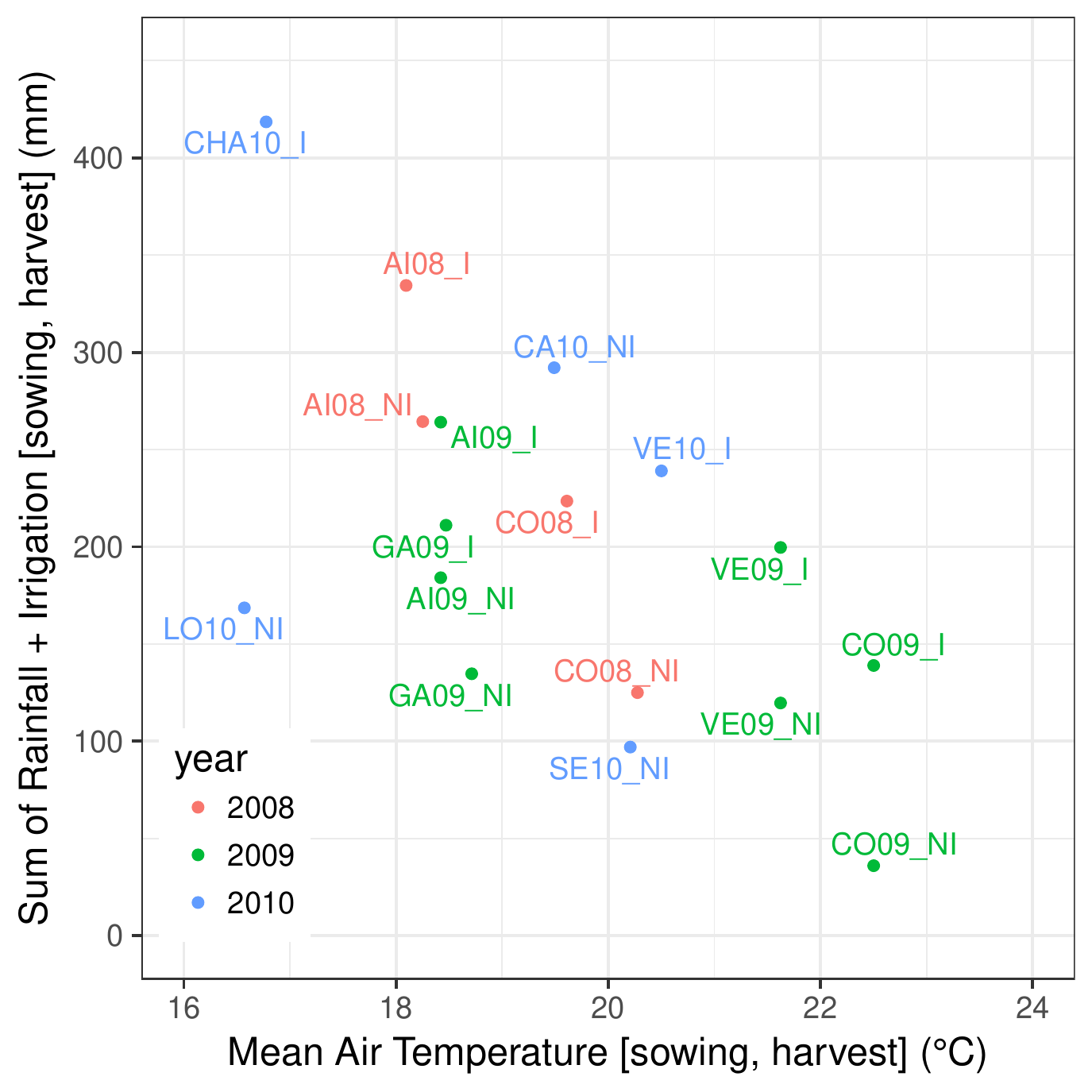}

\begin{quote}
\textbf{Figure S1: Description of the climatic conditions experienced on
the multi-environment trial} Climatic variability on the experimental
network was represented with water (sum of rainfall and precipitations)
and temperature (mean air temperature) indices computed on the cropping
period (sowing-harvest). The 17 locations x year x management
combinations are represented in this space, locations codes are detailed
in table S1.
\end{quote}

\newpage

\includegraphics{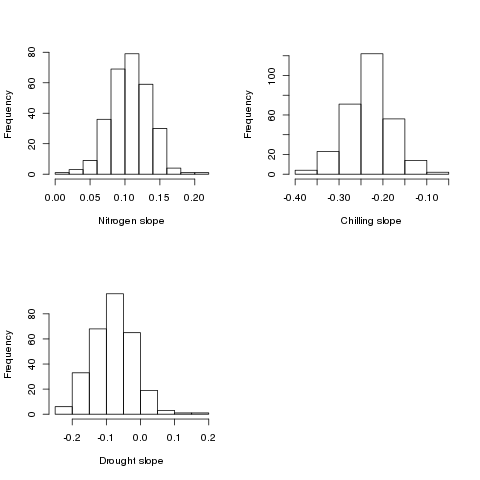}

\begin{quote}
\textbf{Figure S2. Histograms of the plasticity phenotypes for nitrogen
cold and water stresses in the sunflower core collection.}
\end{quote}

\newpage

\subsection*{References}\label{references}
\addcontentsline{toc}{subsection}{References}

\hypertarget{refs}{}
\hypertarget{ref-Acosta1995}{}
Acosta-Gallegos, J., White, J.W., 1995. Phenological plasticity as an
adaptation by common bean to rainfed environments. Crop Science 35,
199--204.
doi:\href{https://doi.org/10.2135/cropsci1995.0011183x003500010037x}{10.2135/cropsci1995.0011183x003500010037x}

\hypertarget{ref-Alonso-Blanco2005}{}
Alonso-Blanco, C., Gomez-Mena, C., Llorente, F., Koornneef, M., Salinas,
J., Martínez-Zapater, J.M., 2005. Genetic and molecular analyses of
natural variation indicate CBF2 as a candidate gene for underlying a
freezing tolerance quantitative trait locus in arabidopsis. Plant
Physiol. 139, 1304--1312.
doi:\href{https://doi.org/10.1104/pp.105.068510}{10.1104/pp.105.068510}

\hypertarget{ref-Andrianasolo2016a}{}
Andrianasolo, F.N., Champolivier, L., Debaeke, P., Maury, P., 2016.
Source and sink indicators for determining nitrogen, plant density and
genotype effects on oil and protein contents in sunflower achenes. Field
Crops Research.
doi:\href{https://doi.org/10.1016/j.fcr.2016.04.010}{10.1016/j.fcr.2016.04.010}

\hypertarget{ref-Blanchet1990}{}
Blanchet, R., Texier, V., Gelfi, N., Viguier, P., 1990. Articulation des
divers processus d'adaptation à la sécheresse et comportements globaux
du tournesol, in: CETIOM (Ed.), Le Tournesol et L'eau : Adaptation à La
Sécheresse, Réponse à L'irrigation. Paris, France, pp. 45--55.

\hypertarget{ref-Brisson2003}{}
Brisson, N., Gary, C., Justes, E., Roche, R., Mary, B., Ripoche, D.,
Zimmer, D., Sierra, J., Bertuzzi, P., Burger, P., Bussière, F.,
Cabidoche, Y.M., Cellier, P., Debaeke, P., Gaudillère, J.P., Hénault,
C., Maraux, F., Seguin, B., Sinoquet, H., 2003. An overview of the crop
model STICS. European Journal of Agronomy 18, 309--332.

\hypertarget{ref-Browning2007}{}
Browning, S.R., Browning, B.L., 2007. Rapid and accurate haplotype
phasing and missing-data inference for whole-genome association studies
by use of localized haplotype clustering. The American Journal of Human
Genetics 81, 1084--1097.
doi:\href{https://doi.org/10.1086/521987}{10.1086/521987}

\hypertarget{ref-Butler2009}{}
Butler, D., Cullis, B., Gilmour, A., Gogel, B., 2009. ASReml-r reference
manual. Queensland Department of Primary Industries, Queensland,
Australia.

\hypertarget{ref-Cabelguenne1999}{}
Cabelguenne, M., Debaeke, P., Bouniols, A., 1999. EPICphase, a version
of the EPIC model simulating the effects of water and nitrogen stress on
biomass and yield, taking account of developmental stages: Validation on
maize, sunflower, sorghum, soybean and winter wheat. Agricultural
Systems 60, 175--196.
doi:\href{https://doi.org/10.1016/s0308-521x(99)00027-x}{10.1016/s0308-521x(99)00027-x}

\hypertarget{ref-Cadic2013}{}
Cadic, E., Coque, M., Vear, F., Grezes-Besset, B., Pauquet, J.,
Piquemal, J., Lippi, Y., Blanchard, P., Romestant, M., Pouilly, N.,
Rengel, D., Gouzy, J., Langlade, N., Mangin, B., Vincourt, P., 2013.
Combined linkage and association mapping of flowering time in sunflower
(helianthus annuus l.). Theor Appl Genet 126, 1337--1356.
doi:\href{https://doi.org/10.1007/s00122-013-2056-2}{10.1007/s00122-013-2056-2}

\hypertarget{ref-Cadic2012}{}
Cadic, E., Debaeke, P., Langlade, N., Grèzes-Besset, B., Pauquet, J.,
Coque, M., André, T., Chatre, S., Casadebaig, P., Mangin, B., Vincourt,
P., 2012. Phenotyping the response of sunflower (helianthus annuus l.)
to drought scenarios in multi-environmental trials for the purpose of
association genetics., in: Plant \& Animal Genomes Xviiii Conference,
San Diego (Usa), 14-18 Jan.

\hypertarget{ref-Casadebaig2008}{}
Casadebaig, P., Debaeke, P., Lecoeur, J., 2008. Thresholds for leaf
expansion and transpiration response to soil water deficit in a range of
sunflower genotypes. European Journal of Agronomy 28, 646--654.
doi:\href{https://doi.org/10.1016/j.eja.2008.02.001}{10.1016/j.eja.2008.02.001}

\hypertarget{ref-Casadebaig2011}{}
Casadebaig, P., Guilioni, L., Lecoeur, J., Christophe, A., Champolivier,
L., Debaeke, P., 2011. SUNFLO, a model to simulate genotype-specific
performance of the sunflower crop in contrasting environments.
Agricultural and Forest Meteorology 151, 163--178.
doi:\href{https://doi.org/10.1016/j.agrformet.2010.09.012}{10.1016/j.agrformet.2010.09.012}

\hypertarget{ref-Chenu2013}{}
Chenu, K., Deihimfard, R., Chapman, S.C., 2013. Large-scale
characterization of drought pattern: A continent-wide modelling approach
applied to the Australian wheatbelt--spatial and temporal trends. New
Phytologist 198, 801--820.
doi:\href{https://doi.org/10.1111/nph.12192}{10.1111/nph.12192}

\hypertarget{ref-Coque2008}{}
Coque, M., Mesnildrey, S., Romestant, M., Grezes-Besset, B., Vear, F.,
Langlade, N., Vincourt, P., 2008. Sunflower line core collections for
association studies and phenomics. Proc. 17th Int Sunflower Conf,
Cordoba (Spain) 725--8.

\hypertarget{ref-Crouzillat1991}{}
Crouzillat, D., Canal, L., Perrault, A., Ledoigt, G., Vear, F., Serieys,
H., 1991. Cytoplasmic male sterility in sunflower: Comparison of
molecular biology and genetic studies. Plant molecular biology 16,
415--426.
doi:\href{https://doi.org/10.1007/bf00023992}{10.1007/bf00023992}

\hypertarget{ref-Debaeke2010}{}
Debaeke, P., Casadebaig, P., Haquin, B., Mestries, E., Palleau, J.-P.,
Salvi, F., 2010. Simulation de la réponse variétale du tournesol à
l'environnement à l'aide du modèle sunflo. Oléagineux, Corps Gras,
Lipides 17, 143--51.
doi:\href{https://doi.org/10.1684/ocl.2010.0308}{10.1684/ocl.2010.0308}

\hypertarget{ref-Debaeke2012a}{}
Debaeke, P., Oosterom, E. van, Justes, E., Champolivier, L., Merrien,
A., Aguirrezabal, L., González-Dugo, V., Massignam, A., Montemurro, F.,
2012. A species-specific critical nitrogen dilution curve for sunflower
(helianthus annuus l.). Field Crops Research 136, 76--84.
doi:\href{https://doi.org/10.1016/j.fcr.2012.07.024}{10.1016/j.fcr.2012.07.024}

\hypertarget{ref-DesMarais2013}{}
Des Marais, D.L., Hernandez, K.M., Juenger, T.E., 2013.
Genotype-by-environment interaction and plasticity: Exploring genomic
responses of plants to the abiotic environment. Annual Review of
Ecology, Evolution, and Systematics 44, 5--29.
doi:\href{https://doi.org/10.1146/annurev-ecolsys-110512-135806}{10.1146/annurev-ecolsys-110512-135806}

\hypertarget{ref-DeWitt2004}{}
DeWitt, T., Langerhans, R., 2004. Integrated solutions to environmental
heterogeneity: Theory of multimoment reaction norms. Phenotypic
Plasticity: Functional and Conceptual Approaches 98--111.

\hypertarget{ref-El-Soda2015}{}
El-Soda, M., Kruijer, W., Malosetti, M., Koornneef, M., Aarts, M.G.M.,
2015. Quantitative trait loci and candidate genes underlying genotype by
environment interaction in the response of arabidopsis thaliana to
drought. Plant Cell and Environment 38, 585--599.
doi:\href{https://doi.org/10.1111/pce.12418}{10.1111/pce.12418}

\hypertarget{ref-Federer1961}{}
Federer, W.T., 1961. Augmented designs with one-way elimination of
heterogeneity. Biometrics 17, 447--473.

\hypertarget{ref-Feng2008}{}
Feng, H., Li, X., Duan, J., Li, H., Liang, H., 2008. Chilling tolerance
of wheat seedlings is related to an enhanced alternative respiratory
pathway. Crop Science 48, 2381.
doi:\href{https://doi.org/10.2135/cropsci2007.04.0232}{10.2135/cropsci2007.04.0232}

\hypertarget{ref-Fukao2011}{}
Fukao, T., Yeung, E., Bailey-Serres, J., 2011. The submergence tolerance
regulator SUB1A mediates crosstalk between submergence and drought
tolerance in rice. Plant Cell 23, 412--427.
doi:\href{https://doi.org/10.1105/tpc.110.080325}{10.1105/tpc.110.080325}

\hypertarget{ref-Gaddour2001}{}
Gaddour, K., Vicente-Carbajosa, J., Lara, P., Isabel-Lamoneda, I., Díaz,
I., Carbonero, P., 2001. A constitutive cystatin-encoding gene from
barley (icy) responds differentially to abiotic stimuli. Plant Mol Biol
45, 599--608.
doi:\href{https://doi.org/10.1023/A:1010697204686}{10.1023/A:1010697204686}

\hypertarget{ref-deGivry2005}{}
Givry, S. de, Bouchez, M., Chabrier, P., Milan, D., Schiex, T., 2005.
CAR(H)(T)AGene: Multipopulation integrated genetic and radiation hybrid
mapping. Bioinformatics 21, 1703--1704.
doi:\href{https://doi.org/10.1093/bioinformatics/bti222}{10.1093/bioinformatics/bti222}

\hypertarget{ref-Groskinsky2015}{}
Großkinsky, D.K., Svensgaard, J., Christensen, S., Roitsch, T., 2015.
Plant phenomics and the need for physiological phenotyping across scales
to narrow the genotype-to-phenotype knowledge gap. Journal of
experimental botany 66, 5429--5440.
doi:\href{https://doi.org/10.1093/jxb/erv345}{10.1093/jxb/erv345}

\hypertarget{ref-Guo2013}{}
Guo, L., Yang, H., Zhang, X., Yang, S., 2013. Lipid transfer protein 3
as a target of MYB96 mediates freezing and drought stress in
arabidopsis. J. Exp. Bot. 64, 1755--1767.
doi:\href{https://doi.org/10.1093/jxb/ert040}{10.1093/jxb/ert040}

\hypertarget{ref-Hincha2001}{}
Hincha, D.K., Neukamm, B., Sror, H.A.M., Sieg, F., Weckwarth, W.,
Rückels, M., Lullien-Pellerin, V., Schröder, W., Schmitt, J.M., 2001.
Cabbage cryoprotectin is a member of the nonspecific plant lipid
transfer protein gene family. Plant Physiol. 125, 835--846.
doi:\href{https://doi.org/10.1104/pp.125.2.835}{10.1104/pp.125.2.835}

\hypertarget{ref-Jones2003}{}
Jones, J.W., Hoogenboom, G., Porter, C., Boote, K., Batchelor, W., Hunt,
L., Wilkens, P., Singh, U., Gijsman, A., Ritchie, J., 2003. The dssat
cropping system model. European journal of agronomy 18, 235--265.
doi:\href{https://doi.org/10.1016/s1161-0301(02)00107-7}{10.1016/s1161-0301(02)00107-7}

\hypertarget{ref-Josse2012}{}
Josse, J., Chavent, M., Liquet, B., Husson, F., 2012. Handling missing
values with regularized iterative multiple correspondence analysis.
Journal of classification 29, 91--116.
doi:\href{https://doi.org/10.1007/s00357-012-9097-0}{10.1007/s00357-012-9097-0}

\hypertarget{ref-Josse2016}{}
Josse, J., Husson, F., 2016. missMDA: A package for handling missing
values in multivariate data analysis. Journal of Statistical Software
70, 1--31.
doi:\href{https://doi.org/10.18637/jss.v070.i01}{10.18637/jss.v070.i01}

\hypertarget{ref-Kane2011}{}
Kane, N.C., Gill, N., Lindhauer, M.G., Bowers, J.E., Bergès, H., Gouzy,
J., Bachlava, E., Langlade, N.B., Lai, Z., Stewart, M., Burke, J.M.,
Vincourt, P., Knapp, S.J., Rieseberg, L.H., 2011. Progress towards a
reference genome for sunflower. Botany 89, 429--437.

\hypertarget{ref-Kang2008}{}
Kang, H.M., Zaitlen, N.A., Wade, C.M., Kirby, A., Heckerman, D., Daly,
M.J., Eskin, E., 2008. Efficient control of population structure in
model organism association mapping. Genetics 178, 1709--1723.
doi:\href{https://doi.org/10.1534/genetics.107.080101}{10.1534/genetics.107.080101}

\hypertarget{ref-Keating2003}{}
Keating, B.A., Carberry, P.S., Hammer, G.L., Probert, M.E., Robertson,
M.J., Holzworth, D., Huth, N.I., Hargreaves, J.N.G., Meinke, H.,
Hochman, Z., 2003. An overview of APSIM, a model designed for farming
systems simulation. European Journal of Agronomy 18, 267--288.
doi:\href{https://doi.org/10.1016/s1161-0301(02)00108-9}{10.1016/s1161-0301(02)00108-9}

\hypertarget{ref-Kesari2012}{}
Kesari, R., Lasky, J.R., Villamor, J.G., Marais, D.L.D., Chen, Y.-J.C.,
Liu, T.-W., Lin, W., Juenger, T.E., Verslues, P.E., 2012.
Intron-mediated alternative splicing of arabidopsis p5cs1 and its
association with natural variation in proline and climate adaptation.
PNAS 109, 9197--9202.
doi:\href{https://doi.org/10.1073/pnas.1203433109}{10.1073/pnas.1203433109}

\hypertarget{ref-Kiani2016}{}
Kiani, M., Gheysari, M., Mostafazadeh-Fard, B., Majidi, M.M., Karchani,
K., Hoogenboom, G., 2016. Effect of the interaction of water and
nitrogen on sunflower under drip irrigation in an arid region.
Agricultural Water Management 171, 162--172.
doi:\href{https://doi.org/10.1016/j.agwat.2016.04.008}{10.1016/j.agwat.2016.04.008}

\hypertarget{ref-Konishi2006}{}
Konishi, M., Sugiyama, M., 2006. A novel plant-specific family gene,
ROOT PRIMORDIUM DEFECTIVE 1, is required for the maintenance of active
cell proliferation. Plant Physiol. 140, 591--602.
doi:\href{https://doi.org/10.1104/pp.105.074724}{10.1104/pp.105.074724}

\hypertarget{ref-Lake2016}{}
Lake, L., Chenu, K., Sadras, V., 2016. Patterns of water stress and
temperature for australian chickpea production. Crop and Pasture
Science. doi:\href{https://doi.org/10.1071/CP15253}{10.1071/CP15253}

\hypertarget{ref-Lecoeur2011}{}
Lecoeur, J., Poiré-Lassus, R., Christophe, A., Pallas, B., Casadebaig,
P., Debaeke, P., Vear, F., Guilioni, L., 2011. Quantifying physiological
determinants of genetic variation for yield potential in sunflower.
SUNFLO: a model-based analysis. Functional Plant Biology 38, 246--259.
doi:\href{https://doi.org/10.1071/fp09189}{10.1071/fp09189}

\hypertarget{ref-Lemaire1997}{}
Lemaire, G., Meynard, J., 1997. Use of the nitrogen nutrition index for
the analysis of agronomical data, in: Lemaire, G. (Ed.), Diagnosis of
the Nitrogen Status in Crops. Ed. G. Lemaire. Springer-Verlag, Berlin,
pp. 45--55.
doi:\href{https://doi.org/10.1007/978-3-642-60684-7_2}{10.1007/978-3-642-60684-7\_2}

\hypertarget{ref-Li2005}{}
Li, J., Ji, L., 2005. Adjusting multiple testing in multilocus analyses
using the eigenvalues of a correlation matrix. Heredity 95, 221--227.
doi:\href{https://doi.org/10.1038/sj.hdy.6800717}{10.1038/sj.hdy.6800717}

\hypertarget{ref-Liu2015}{}
Liu, F., Zhang, X., Lu, C., Zeng, X., Li, Y., Fu, D., Wu, G., 2015.
Non-specific lipid transfer proteins in plants: Presenting new advances
and an integrated functional analysis. J. Exp. Bot. 66, 5663--5681.
doi:\href{https://doi.org/10.1093/jxb/erv313}{10.1093/jxb/erv313}

\hypertarget{ref-Lobell2008}{}
Lobell, D.B., Burke, M.B., Tebaldi, C., Mastrandrea, M.D., Falcon, W.P.,
Naylor, R.L., 2008. Prioritizing climate change adaptation needs for
food security in 2030. Science 319, 607--610.
doi:\href{https://doi.org/10.1126/science.1152339}{10.1126/science.1152339}

\hypertarget{ref-Maffia2003}{}
Maffia, M., Rizzello, A., Acierno, R., Verri, T., Rollo, M., Danieli,
A., Döring, F., Daniel, H., Storelli, C., 2003. Characterisation of
intestinal peptide transporter of the antarctic haemoglobinless teleost
chionodraco hamatus. Journal of Experimental Biology 206, 705--714.
doi:\href{https://doi.org/10.1242/jeb.00145}{10.1242/jeb.00145}

\hypertarget{ref-Mahalingam2015}{}
Mahalingam, R., 2015. Consideration of combined stress: A crucial
paradigm for improving multiple stress tolerance in plants, in:
Mahalingam, R. (Ed.), Combined Stresses in Plants. Springer
International Publishing, pp. 1--25.
doi:\href{https://doi.org/10.1007/978-3-319-07899-1_1}{10.1007/978-3-319-07899-1\_1}

\hypertarget{ref-Mangin2012}{}
Mangin, B., Siberchicot, A., Nicolas, S., Doligez, A., This, P.,
Cierco-Ayrolles, C., 2012. Novel measures of linkage disequilibrium that
correct the bias due to population structure and relatedness. Heredity
108, 285--291.
doi:\href{https://doi.org/10.1038/hdy.2011.73}{10.1038/hdy.2011.73}

\hypertarget{ref-Massonneau2005}{}
Massonneau, A., Condamine, P., Wisniewski, J.P., Zivy, M., Rogowsky,
P.M., 2005. Maize cystatins respond to developmental cues, cold stress
and drought. Biochimica Et Biophysica Acta-Gene Structure and Expression
1729, 186--199.
doi:\href{https://doi.org/10.1016/j.bbaexp.2005.05.004}{10.1016/j.bbaexp.2005.05.004}

\hypertarget{ref-McKay2008}{}
McKay, J., Richards, J., Nemali, K., Sen, S., Mitchell-Olds, T., Boles,
S., Stahl, E., Wayne, T., Juenger, T., 2008. Genetics of drought
adaptation in arabidopsis thaliana II. QTL analysis of a new mapping
population, kas-1 x tsu-1. Evolution 62, 3014.
doi:\href{https://doi.org/10.1111/j.1558-5646.2008.00474.x}{10.1111/j.1558-5646.2008.00474.x}

\hypertarget{ref-Monteith1994}{}
Monteith, J.L., 1994. Validity of the correlation between intercepted
radiation and biomass. Agricultural and Forest Meteorology 68, 213--220.
doi:\href{https://doi.org/10.1016/0168-1923(94)90037-x}{10.1016/0168-1923(94)90037-x}

\hypertarget{ref-Monteith1977}{}
Monteith, J.L., 1977. Climate and the Efficiency of Crop Production in
Britain. Philosophical Transactions of the Royal Society of London.
Series B, Biological Sciences 281, 277--294.
doi:\href{https://doi.org/10.2307/2402584}{10.2307/2402584}

\hypertarget{ref-Nicotra2010}{}
Nicotra, A.B., Atkin, O.K., Bonser, S.P., Davidson, A.M., Finnegan,
E.J., Mathesius, U., Poot, P., Purugganan, M.D., Richards, C.L.,
Valladares, F., Kleunen, M. van, 2010. Plant phenotypic plasticity in a
changing climate. Trends in Plant Science 15, 684--692.
doi:\href{https://doi.org/10.1016/j.tplants.2010.09.008}{10.1016/j.tplants.2010.09.008}

\hypertarget{ref-Pagnussat2009}{}
Pagnussat, L.A., Lombardo, C., Regente, M., Pinedo, M., Martin, M.,
Canal, L. de la, 2009. Unexpected localization of a lipid transfer
protein in germinating sunflower seeds. Journal of Plant Physiology 166,
797--806.
doi:\href{https://doi.org/10.1016/j.jplph.2008.11.005}{10.1016/j.jplph.2008.11.005}

\hypertarget{ref-Pagnussat2015}{}
Pagnussat, L.A., Oyarburo, N., Cimmino, C., Pinedo, M.L., Canal, L. de
la, 2015. On the role of a lipid-transfer protein. arabidopsis ltp3
mutant is compromised in germination and seedling growth. Plant
Signaling \& Behavior 10, e1105417.
doi:\href{https://doi.org/10.1080/15592324.2015.1105417}{10.1080/15592324.2015.1105417}

\hypertarget{ref-Prins2008}{}
Prins, A., Van Heerden, P.D.R., Olmos, E., Kunert, K.J., Foyer, C.H.,
2008. Cysteine proteinases regulate chloroplast protein content and
composition in tobacco leaves: A model for dynamic interactions with
ribulose-1,5-bisphosphate carboxylase/oxygenase (rubisco) vesicular
bodies. Journal of Experimental Botany 59, 1935--1950.
doi:\href{https://doi.org/10.1093/jxb/ern086}{10.1093/jxb/ern086}

\hypertarget{ref-Quint2009}{}
Quint, M., Barkawi, L.S., Fan, K.-T., Cohen, J.D., Gray, W.M., 2009.
Arabidopsis IAR4 modulates auxin response by regulating auxin
homeostasis. Plant Physiol. 150, 748--758.
doi:\href{https://doi.org/10.1104/pp.109.136671}{10.1104/pp.109.136671}

\hypertarget{ref-R2014}{}
R Core Team, 2014. R: A language and environment for statistical
computing. R Foundation for Statistical Computing, Vienna, Austria.

\hypertarget{ref-Ren2010}{}
Ren, Z., Zheng, Z., Chinnusamy, V., Zhu, J., Cui, X., Iida, K., Zhu,
J.-K., 2010. RAS1, a quantitative trait locus for salt tolerance and ABA
sensitivity in arabidopsis. PNAS 107, 5669--5674.
doi:\href{https://doi.org/10.1073/pnas.0910798107}{10.1073/pnas.0910798107}

\hypertarget{ref-Ribas2000}{}
Ribas-Carbo, M., Aroca, R., Gonzàlez-Meler, M.A., Irigoyen, J.J.,
Sánchez-Díaz, M., 2000. The electron partitioning between the cytochrome
and alternative respiratory pathways during chilling recovery in two
cultivars of maize differing in chilling sensitivity. Plant Physiol.
122, 199--204.
doi:\href{https://doi.org/10.1104/pp.122.1.199}{10.1104/pp.122.1.199}

\hypertarget{ref-Rizzello2013}{}
Rizzello, A., Romano, A., Kottra, G., Acierno, R., Storelli, C., Verri,
T., Daniel, H., Maffia, M., 2013. Protein cold adaptation strategy via a
unique seven-amino acid domain in the icefish (chionodraco hamatus)
PEPT1 transporter. PNAS 110, 7068--7073.
doi:\href{https://doi.org/10.1073/pnas.1220417110}{10.1073/pnas.1220417110}

\hypertarget{ref-Rosenzweig1994}{}
Rosenzweig, C., Parry, M.L., others, 1994. Potential impact of climate
change on world food supply. Nature 367, 133--138.
doi:\href{https://doi.org/10.1038/367133a0}{10.1038/367133a0}

\hypertarget{ref-Sadras2016}{}
Sadras, V.O., Denison, R.F., 2016. Neither crop genetics nor crop
management can be optimised. Field Crops Research.
doi:\href{https://doi.org/10.1016/j.fcr.2016.01.015}{10.1016/j.fcr.2016.01.015}

\hypertarget{ref-Sadras2009a}{}
Sadras, V.O., Reynolds, M.P., De la Vega, A., Petrie, P.R., Robinson,
R., 2009. Phenotypic plasticity of yield and phenology in wheat,
sunflower and grapevine. Field Crops Research 110, 242--250.
doi:\href{https://doi.org/10.1016/j.fcr.2008.09.004}{10.1016/j.fcr.2008.09.004}

\hypertarget{ref-Sambatti2007}{}
Sambatti, J., Caylor, K.K., 2007. When is breeding for drought tolerance
optimal if drought is random? New Phytologist 175, 70--80.
doi:\href{https://doi.org/10.1111/j.1469-8137.2007.02067.x}{10.1111/j.1469-8137.2007.02067.x}

\hypertarget{ref-Segura2012}{}
Segura, V., Vilhjálmsson, B.J., Platt, A., Korte, A., Seren, Ü., Long,
Q., Nordborg, M., 2012. An efficient multi-locus mixed-model approach
for genome-wide association studies in structured populations. Nature
genetics 44, 825--830.
doi:\href{https://doi.org/10.1038/ng.2314}{10.1038/ng.2314}

\hypertarget{ref-Self1987}{}
Self, S.G., Liang, K.-Y., 1987. Asymptotic properties of maximum
likelihood estimators and likelihood ratio tests under nonstandard
conditions. Journal of the American Statistical Association 82,
605--610.
doi:\href{https://doi.org/10.1080/01621459.1987.10478472}{10.1080/01621459.1987.10478472}

\hypertarget{ref-Serieys1984}{}
Serieys, H., 1984. Wild helianthus species, a potential source of
androsterilities, in: Second Eucarpia Meeting on the Sunflower,
Leningrad. pp. 10--14.

\hypertarget{ref-Sosnowski2012}{}
Sosnowski, O., Charcosset, A., Joets, J., 2012. BioMercator v3: An
upgrade of genetic map compilation and quantitative trait loci
meta-analysis algorithms. Bioinformatics 28, 2082--2083.
doi:\href{https://doi.org/10.1093/bioinformatics/bts313}{10.1093/bioinformatics/bts313}

\hypertarget{ref-Talanova2012}{}
Talanova, V.V., Titov, A.F., Topchieva, L.V., Frolova, S.A., 2012.
Effects of abscisic acid treatment on the expression of cysteine
proteinase gene and enzyme inhibitor during wheat cold adaptation.
Russian Journal of Plant Physiology 59, 581--585.
doi:\href{https://doi.org/10.1134/S1021443712040140}{10.1134/S1021443712040140}

\hypertarget{ref-VanEeuwijk2010}{}
Van Eeuwijk, F., Bink, M., Chenu, K., Chapman, S., 2010. Detection and
use of qtl for complex traits in multiple environments. Current opinion
in plant biology 13, 193--205.
doi:\href{https://doi.org/10.1016/j.pbi.2010.01.001}{10.1016/j.pbi.2010.01.001}

\hypertarget{ref-Via1985}{}
Via, S., Lande, R., 1985. Genotype-environment interaction and the
evolution of phenotypic plasticity. Evolution 505--522.
doi:\href{https://doi.org/10.2307/2408649}{10.2307/2408649}

\hypertarget{ref-Vile2012}{}
Vile, D., Pervent, M., Belluau, M., Vasseur, F., Bresson, J., Muller,
B., Granier, C., Simonneau, T., 2012. Arabidopsis growth under prolonged
high temperature and water deficit: Independent or interactive effects?
Plant, Cell \& Environment.
doi:\href{https://doi.org/10.1111/j.1365-3040.2011.02445.x}{10.1111/j.1365-3040.2011.02445.x}

\hypertarget{ref-Villalobos1996}{}
Villalobos, F., Hall, A., Ritchie, J., Orgaz, F., 1996. OILCROP-SUN: A
development, growth and yield model of the sunflower crop. Agronomy
Journal 88, 403--415.
doi:\href{https://doi.org/10.2134/agronj1996.00021962008800030008x}{10.2134/agronj1996.00021962008800030008x}

\hypertarget{ref-Villanova2011}{}
Villanova, L., Smith-Miles, K., Hyndman., R.J., 2011. EMMA: Evolutionary
Model-based Multiresponse Approach.

\hypertarget{ref-Yu2006}{}
Yu, J., Pressoir, G., Briggs, W.H., Bi, I.V., Yamasaki, M., Doebley,
J.F., McMullen, M.D., Gaut, B.S., Nielsen, D.M., Holland, J.B., others,
2006. A unified mixed-model method for association mapping that accounts
for multiple levels of relatedness. Nature genetics 38, 203--208.

\hypertarget{ref-Zhang2008}{}
Zhang, X., Liu, S., Takano, T., 2008. Two cysteine proteinase inhibitors
from arabidopsis thaliana, AtCYSa and AtCYSb, increasing the salt,
drought, oxidation and cold tolerance. Plant Mol Biol 68, 131--143.
doi:\href{https://doi.org/10.1007/s11103-008-9357-x}{10.1007/s11103-008-9357-x}

\end{document}